\documentclass[12pt]{article}
\pdfoutput=1
\usepackage{jheppreprint}
\usepackage{tikz}
\usepackage{bm}
\usepackage{amssymb}
\title{Lorentz violating kinematics: \newline Threshold theorems}
\author{Valentina Baccetti,  Kyle Tate, \textmd{and} Matt Visser}
\affiliation{School of Mathematics, Statistics, and Operations Research, \\
Victoria University of Wellington, PO Box 600, Wellington 6140, New Zealand}
\emailAdd{valentina.baccetti@msor.vuw.ac.nz}
\emailAdd{kyle.tate@msor.vuw.ac.nz}
\emailAdd{matt.visser@msor.vuw.ac.nz}
\abstract{Recent tentative experimental indications, and the subsequent theoretical speculations, regarding possible violations of Lorentz invariance have attracted a vast amount of attention. 
An important  technical issue that considerably complicates detailed calculations in any such scenario, is that once one violates Lorentz invariance the analysis of thresholds in both scattering and decay processes becomes  extremely subtle, with many new and naively unexpected effects. In the current article we develop several extremely general threshold theorems that depend only on the existence of some energy momentum relation $E(\bm{p})$, eschewing even assumptions of isotropy or monotonicity. We shall argue that there are physically interesting situations where such a level of generality is called for, and that existing (partial) results in the literature make unnecessary technical assumptions. Even in this most general of settings, we show that at threshold all final state particles move with the same 3-velocity, while initial state particles must have 3-velocities parallel/anti-parallel to the final state particles. In contrast the various 3-momenta can behave in a complicated and counter-intuitive manner.

\bigskip
\noindent
28 November 2011; \LaTeX-ed \today
}
\begin{document}
\maketitle
\def\x{\bm{x}}
\def\p{\bm{p}}
\def\0{\bm{0}}
\def\n{\bm{n}}
\def\vv{\bm{v}}
\def\u{\bm{u}}
\def\w{\bm{w}}
\def\bpi{\bm{\pi}}
\def\blambda{\bm{\lambda}}
\def\P{{\mathcal{P}}}
\def\C{{\mathcal{C}}}
\def\E{{\mathcal{E}}}
\def\R{{\mathbb{R}}}
\newpage
\section{Introduction}
\label{S:intro}

The OPERA collaboration's announcement of tentative but statistically significant indications of ``faster than light'' neutrinos~\cite{Opera:2011} (see also earlier more tentative results from the MINOS collaboration~\cite{Minos:2007}) has ignited a firestorm of theoretical speculation. 
Over 150 theoretical articles have been generated in some $9{1\over2}$ weeks.  
Notable contributions include~\cite{AmelinoCamelia:2011dx, Giudice:2011mm, Cohen:2011hx, Dvali:2011mn, Alexandre:2011bu, Cacciapaglia:2011ax, Bi:2011nd, Klinkhamer:2011mf, Gubser:2011mp, Kehagias:2011cb, Wang:2011sz, Saridakis:2011eq, Winter:2011zf, Alexandre:2011kr, Klinkhamer:2011iz, Cowsik:2011wv, Maccione:2011fr, Dass:2011yj, Carmona:2011zg}. 
In addition to these very  recent efforts, it is important to recognize that there is an older and extensive literature placing significant experimental and observational bounds on any possible violation of Lorentz invariance. 
See for instance work by Coleman and Glashow~\cite{Coleman:1997xq, Coleman:1998ti}, 
Jacobson and collaborators~\cite{Liberati:2001cr, Jacobson:2001tu,  Jacobson:2002hd, Mattingly:2002ba, Jacobson:2002ye, Jacobson:2003ty, Jacobson:2003bn, Jacobson:2004qt, Jacobson:2004rj, Jacobson:2005bg}, and especially the Living Review by Mattingly~\cite{Mattingly:2005re}. 
Other theoretical frameworks for characterizing possible violations of Lorentz invariance include those of Nielsen and collaborators~\cite{NBI-HE-78-10, NBI-HE-82-42, NBI-HE-82-30, 187008, NBI-HE-82-9},  Kostelecky and collaborators~\cite{Colladay:1998fq, Kostelecky:1988zi, Kostelecky:2003fs, Kostelecky:2000mm, Kostelecky:2002hh, Kostelecky:2003cr, Kostelecky:1999mr, Kostelecky:2001mb, Bear:2000cd}, and the flat-space non-gravity framework developed by Anselmi~\cite{Anselmi:2011zz, Anselmi:2011bp, Anselmi:2007zz, Anselmi:2011ae, Anselmi:2011ac, Anselmi:2010zh, Anselmi:2009ng, Anselmi:2009vz, Anselmi:2008bt, Anselmi:2008bs, Anselmi:2008bq, Anselmi:2008ry, Anselmi:2007ri}.  
Additionally, the Ho\v{r}ava gravity framework~\cite{Horava:2009uw} naturally includes Lorentz violation~\cite{Visser:2009fg, Visser:2009ys, Sotiriou:2009bx, Sotiriou:2009gy, Weinfurtner:2010hz, Visser:2011mf}, though quantitatively the relationship between Ho\v{r}ava gravity and the OPERA results seems somewhat strained~\cite{Saridakis:2011eq}. 
In this article we shall not directly address the phenomenology of the OPERA--MINOS results. Instead we shall prove some very general theorems on Lorentz violating processes, theorems that should be borne in mind whenever one is attempting to step outside the framework of standard special and general relativity.

One of the key results in the literature devoted to possible violations of Lorentz invariance is that the normal intuition one develops regarding threshold phenomena requires significant modification. See for instance the articles by Coleman and Glashow~\cite{Coleman:1997xq, Coleman:1998ti},  and Jacobson, Liberati, and Mattingly~\cite{Mattingly:2002ba, Jacobson:2002hd} --- and the more recent follow-ups by Cohen and Glashow~\cite{Cohen:2011hx}, and Liberati, Mattingly, and Maccione~\cite{Maccione:2011fr}, focussing specifically on the OPERA results.
In this article we will generalize the analysis of threshold phenomena presented in those articles. We shall consider both single-particle decay processes, and two-particle scattering processes (possibly inelastic), taking care to make an absolute minimum of technical assumptions --- thus greatly generalizing previous analyses. Specifically:
\begin{itemize}
\item 
We will explicitly assume a normal spacetime manifold based on $\R^4$. 
\\
(This excludes, for instance, both non-commutative spacetimes and certain versions of DSR~\cite{Judes:2002bw, Liberati:2004ju}. This is a purely pragmatic decision based on the fact that we want to be able to say something reasonably concrete.)
\item 
We shall explicitly assume  conservation of both energy and momentum.\\
(This is again a purely pragmatic decision based on the fact that we want to be able to say something reasonably concrete.)

\item
Furthermore we shall explicitly assume a Hamiltonian/Lagrangian framework, so that in view of Noether's theorem (combined with energy-momentum conservation) we are working in a homogeneous spacetime. Specifically, we assume the free-particle energy to be some function of the 3-momentum, $E(\p)$, and that this can be related to a 3-velocity via Hamilton's equations
\begin{equation}
\dot\x = {\partial E\over\partial\p}; \qquad  \dot \p = 0. 
\end{equation}
(This is again a purely pragmatic decision based on the fact that we want to have a sufficiently well-defined framework in which to be able to say something reasonably concrete.)

\item
We shall eschew any \emph{particular} functional form for $E(\p)$, though we will impose smoothness and differentiability constraints as needed.

\item
We shall explicitly \emph{not} assume isotropy.
\begin{itemize}
\item 
Even if physics happens to be isotropic in the preferred (``aether'') frame  implicit in many specific Lorentz violating theories, there is no particular reason to assume isotropy of the energy-momentum relation in generic inertial frames.  And since we do not necessarily know what the observer's  3-velocity is with respect to the preferred frame, it is more useful to develop threshold analysis for generic observers in an explicitly observer-dependent manner.
\item
Even as early as the 1980's attempts were made to take lattice physics seriously as a physical cutoff --- with proton decay taking place with outgoing decay products preferentially aligned along the principal axes of the universe.  In such a situation one would not have isotropy even in the preferred frame. 
\item 
Many of the ``analogue spacetime'' models permit energy-momentum relations that have odd and possibly anisotropic behaviors at ultra-high energies~\cite{Barcelo:2005fc, Visser:1997ux}.
\item
Consider a generic quasiparticle propagating in a generic atomic lattice.  (For instance, a conduction-band dressed electron.) The band structure will typically not be isotropic, even in the rest frame of the lattice.  So if you are a condensed matter physicist, you will have no choice, you will simply have to acknowledge that non-isotropy of the energy-momentum relation is quite common, which will unavoidably influence your ability to analyze reaction thresholds.  
 \end{itemize}

\item
We shall also be extremely cautious concerning ``monotonicity'' assumptions --- carefully formulating an appropriate  concept of monotonicity,  and carefully analyzing what can and cannot be extracted from such an assumption.

\end{itemize}
Because of the generality of these assumptions, our results will have considerably wider validity than the results currently extant in the literature. Even in this most general of frameworks, several rigorous theorems can be extracted. We shall show that at threshold all final state particles move with the same 3-velocity, while initial state particles must have 3-velocities parallel/anti-parallel to the final state particles. In contrast the various 3-momenta can behave in quite complicated and counter-intuitive fashion, and the 3-momenta need not even be collinear.

\section{General background} \label{sec: general}

In the Coleman--Glashow analysis~\cite{Coleman:1998ti},  most of the discussion is explicitly limited to the rather special case of single-particle decay processes where the initial and final state particles $\left( i\in\{1,2,\dots,n\} \right)$ all have energies of the form
\begin{equation}
E_i(\p_i)= \sqrt{ E_{i,0}^2 + ||\p_i||^2 c_i^2}.
\label{E:Coleman}
\end{equation}
Here the ``speed of light'' can be particle dependent. (See especially equation (2.19) in reference~\cite{Coleman:1998ti}. A similar assumption is implicitly made in~\cite{Coleman:1997xq}.)
But it is reasonably clear that much of the discussion of thresholds in~\cite{Coleman:1998ti} would work for any generic $E_i(\p_i)$. 
In contrast, in the OPERA-related analysis of~\cite{Cohen:2011hx}, this specific choice of energy-momentum relation is implicit, not explicit, but is absolutely essential to that discussion --- see~\cite{Maccione:2011fr} for a generalization.

The Mattingly, Jacobson, and Liberati threshold analysis~\cite{Mattingly:2002ba} focusses on energy-momentum relations that are (even in their most general setting) taken to be both isotropic and monotonic. Specific examples are taken to be of the form
\begin{equation}
E_i(\p_i)= \sqrt{ E_{i,0}^2 + ||\p_i||^2 c^2 + \eta_i \; {(||\p_i||\,c)^n E_*^{2-n}}} .
\label{E:Mattingly}
\end{equation}
Here $\eta_i$ is a dimensionless parameter and $E_*$ is an energy scale characterizing the deviations from Lorentz invariance. 
The special case $n=2$ corresponds to the Coleman--Glashow energy-momentum relation.  If $\eta_i<0$ and $n>2$ then these energy-momentum relations can in principle exhibit a maximum --- the energy ``saturates'' --- this is a specific example of a much more general phenomenon:
\begin{itemize}

\item 
In lattice QFT regularizations energy-momentum relations are typically of the form
\begin{equation}
E(\p) =  
\sqrt{ E_0^2 + \left({\hbar c\over a}\right)^2 \; \left\{
\sin^2\left({\p_x a \over \hbar}\right) + \sin^2\left({\p_y a \over \hbar}\right)  +  \sin^2\left({\p_z a \over \hbar}\right) \right\}   },
\label{E:lattice}
\end{equation}
or, (for ``massless'' particles), 
\begin{equation}
E(\p) =  {\hbar c\over a} \;  
\sqrt{
\sin^2\left({\p_x a \over \hbar}\right) + \sin^2\left({\p_y a \over \hbar}\right)  +  \sin^2\left({\p_z a \over \hbar}\right)   },
\label{E:lattice2}
\end{equation}
and typically exhibit a maximum energy $E_\mathrm{max}=\sqrt{E_0^2+3 \hbar^2 c^2/a^2}$ in terms of the lattice spacing $a$. 
(One usually considers ``small'' momenta, $||\p||a/\hbar\ll1$, where Lorentz invariance is approximately recovered. Herein the focus will be on deviations from Lorentz invariance.)

\item 
Qualitatively similar effects occur  for quasiparticles propagating through atomic lattices --- momentum space [the first Brillouin zone] is now compact, and so (assuming continuity) the energy will be bounded by some maximum.

\item
In DSR-inspired models, insofar as they can be incorporated into the current framework,  one often has individual particle energies saturating at or around the Planck energy~\cite{Judes:2002bw, Liberati:2004ju}.

\item
For a specific example (physically unmotivated but mathematically tractable) of saturation behaviour one might take
\begin{equation}
E(\p) =  \sqrt{E_0^2 + E_*^2 \; \tanh\left( {||\p||^2 c^2\over E_*^2} \right) }.
\label{E:tanh}
\end{equation}
At low momentum this is approximately Lorentz invariant
\begin{equation}
E(\p) \approx  \sqrt{E_0^2 + ||\p||^2 c^2 +\mathcal{O}(||\p||^4)},
\end{equation}
but at high momentum it exponentially saturates
\begin{equation}
E(\p) \approx  \sqrt{E_0^2 + E_*^2 +\mathcal{O}(\exp(-2||\p||^2 c^2/E_*^2))}.
\end{equation}

\item
Neither the usual Lorentz invariant energy-momentum relation, nor even the Galilean invariant energy-momentum relation, saturate with a maximum energy.
For that matter, neither does the Coleman--Glashow energy-momentum relation (\ref{E:Coleman}) saturate.

\end{itemize}
Since the whole point of this article is to generalize threshold analysis as far as possible to situations where Lorentz invariance is broken, we will for generality entertain the possibility of energy-momentum relations that saturate to some maximum energy --- with the understanding that this maximum energy might, in specific situations, be infinite.  
With this general framework in place we are now ready to begin detailed analysis. 

\section{Cautionary comments}
\label{S:caution}
\def\sech{\mathrm{sech}}

Some cautionary comments are in order:
\begin{itemize}
\item For the lattice-like energy-momentum relation of equation (\ref{E:lattice}) we have
\begin{equation}
\vv = {\partial E\over\partial\p} = 
{\hbar c\over 2E a} 
\left( \sin\left({2\p_xa\over\hbar}\right),   \sin\left({2\p_ya\over\hbar}\right),   \sin\left({2\p_za\over\hbar}\right)   \right).
\end{equation}
The key point is that 3-velocity $\vv$ and 3-momentum $\p$ need not be parallel. Additionally $\vv$ can exhibit non-trivial zeros for non-zero momentum $\p$, and even once one specifies a particular particle the inverse function $\p(\vv)$ can easily be (and typically is) multivalued. Such phenomena are not limited to the specific energy-momentum relation of equation (\ref{E:lattice}), but rather are generic to any quasiparticle propagating through a regular lattice (for example, a conduction-band dressed electron).

\item
Non uniqueness of the inverse function  $\p(\vv)$ is also generic for (higher than quadratic) polynomial or rational polynomial energy-momentum relations --- it is the unique invertability of the Lorentz invariant energy-momentum relation $E = \sqrt{E_*^2+||\p||^2 c^2}$ that is \emph{non generic} in this regard.
(Details depend on the precise values of the coefficients as the potential for multi-valued behaviour depends on the root structure.)

\item
 For the \emph{tanh}-like energy-momentum relation of equation (\ref{E:tanh}) we have
\begin{equation}
\vv = {\partial E\over\partial\p} = 
c \; \sech^2\left({||\p||^2 c^2\over E_*^2}\right) {\p c\over E}.
\end{equation}
While  3-velocity $\vv$ and 3-momentum $\p$ are now parallel, zero 3-velocity can correspond either to zero 3-momentum \emph{or} to infinite 3-momentum (with finite energy $\sqrt{E_0^2+E_*^2}$).  Low-velocity physics can thus be grossly misleading --- two particles with the same 3-velocity may have wildly differing 3-momenta. Such phenomena are not limited to the specific energy-momentum relation of equation (\ref{E:tanh}), but rather are generic to any situation where the energy saturates as a function of 3-momentum.
Note in particular that the energy-momentum relation of equation (\ref{E:tanh}) is monotonic --- monotonicity is not enough to prevent this sort of behaviour. Similar behaviour also occurs whenever $\lim_{||\p||\to\infty} \partial E/\partial \p \to \0$, corresponding to a sub-linear asymptotic growth in the energy-momentum relation.

\item
Other unusual possibilities include energy minima occurring at non-zero 3-momen\-tum, (by definition an energy minimum must always occur at zero 3-velocity).  Let $\n$ be an arbitrary unit vector and consider for instance
\begin{equation}
E = \sqrt{ E_0^2 + ||\p||^2 c^2 + k_4 (\p\cdot \n)^4 + k_6 (\p\cdot\n)^6 }.
\label{E:anisotropic}
\end{equation}
This energy-momentum relation is not only anisotropic, but by taking $k_4<0$ and $k_6>0$ one can arrange for a global minimum energy at some $\p_\mathrm{min} = p_\mathrm{min} \n \neq \0$.  The 3-velocity is
\begin{equation}
\vv = { \p c^2 + 2 k_4 (\p\cdot \n)^3 \n + 3 k_6 (\p\cdot \n)^5 \n \over E},
\end{equation}
and $\p$ and $\vv$ are generally not collinear (unless one happens to be considering motion parallel or perpendicular to the preferred axis $\n$.) 

\item
As a final pedagogical example consider the isotropic energy-momentum relation
\begin{equation}
E = \sqrt{ E_0^2 + ||\p||^2 c^2 + k_4 ||\p||^4 + k_6 ||\p||^6 }
\end{equation}
By taking $k_4<0$ and $k_6>0$ one can arrange for a global minimum energy at some finite $||\p_\mathrm{min}|| = p_\mathrm{min} \neq 0$.  The 3-velocity is
\begin{equation}
\vv = {  ||\p|| c^2 + 2 k_4 ||\p||^3 + 3 k_6 ||\p||^5  \over E}  \; \hat\p.
\end{equation}
In this situation $\p$ and $\vv$ are generally collinear, but whenever there is a global energy minimum at some finite $||\p_\mathrm{min}|| = p_\mathrm{min} \neq 0$ there will also be a non-empty range of momenta for which $\p$ and $\vv$ are \emph{anti-parallel}.

\end{itemize}
These are merely five specific examples of the unusual behaviour one might potentially encounter, and the types of issues we shall potentially need to consider in our analysis.

\section{Decay thresholds}
\label{S:decay}

Consider the decay process
\begin{equation}
X_0 \to  X_1 + X_2 + \dots X_n,
\label{E:decay}
\end{equation}
where for each individual particle we have the 4-momenta
\begin{equation}
P_i = (E_i(\p), \p_i).
\end{equation}
We shall now study the kinematics of this decay process. 

\subsection{Kinematically allowed region} 
\label{S:kinematic}

Let us define
\begin{equation}
E_\mathrm{out}(\p_1,\p_2,\dots,\p_n) = \sum_{i=1}^n E_i(\p_i),
\end{equation}
and
\begin{equation}
\P_\mathrm{out}(\p_0) = \left\{ (\p_1, \p_2, \dots, \p_n) : \sum_{i=1}^n \p_i = \p_0\right\}. 
\end{equation}
Then $\P_\mathrm{out}(\p_0)$ is a collection of three individually connected hyperplanes in $\R^n$, one hyperplane for each Cartesian component of $\p_0$,  corresponding to the set of all possible outgoing 3-momenta for fixed total 3-momentum $\p_0$. Thus $\P_\mathrm{out}(\p_0)$ is a $3n-3$ dimensional plane (affine subspace) of co-dimension 3 in $\R^{3n}$, and is both convex and connected as a subset of $\R^{3n}$. 
But the individual $E_i(\p_i)$ are by assumption differentiable and continuous, so $E_\mathrm{out}(\p_1,\p_2,\dots,\p_n)$ is also differentiable and continuous. In particular, since $\P_\mathrm{out}(\p_0)$ is connected, this implies the image $E_\mathrm{out}(\P_\mathrm{out}(\p_0))$ is a connected interval in $\R$. 

For specified initial 3-momentum $\p_0$, the decay process (\ref{E:decay}) is kinematically allowed if and only if
\begin{equation}
E_0(\p_0) \in E_\mathrm{out}(\P_\mathrm{out}(\p_0)).
\end{equation}
That is, the decay is allowed if and only if among the set of all possible output 3-momenta $\{\p_1,\p_2,\dots,\p_n\}$ that conserve total 3-momentum, there is at least one configuration that also conserves total energy. 
We could also phrase the kinematically allowed region in terms of an allowable set of output momenta by considering the inverse image
\begin{equation}
E_\mathrm{out}^{-1}\left(E_0(\p_0)\right)  \cap  \P_\mathrm{out}(\p_0),
\end{equation}
a set which, for given $\p_0$, may or may not be empty.  Alternatively one can ask the question
\begin{equation}
\p_0 \in E_0^{-1}\left(E_\mathrm{out}(\P_\mathrm{out}(\p_0))\right)?
\end{equation}
But these approaches in terms of inverse images can be somewhat clumsy.

As a more practical way to better characterize the kinematically allowed region, it is useful to introduce the two quantities
\begin{equation}
E_\mathrm{min}(\p_0) 
=   \min E_\mathrm{out}(\P_\mathrm{out}(\p_0))    
=    \min \left\{ \sum_{i=1}^n E_i(\p_i) : \sum_{i=1}^n \p_i = \p_0 \right\}, 
\end{equation}
and
\begin{equation}
E_\mathrm{max}(\p_0) 
=   \max E_\mathrm{out}(\P_\mathrm{out}(\p_0))    
=    \max \left\{ \sum_{i=1}^n E_i(\p_i) : \sum_{i=1}^n \p_i = \p_0 \right\}.
\end{equation}
A more technically precise statement would use the concepts of \emph{supremum} and \emph{infimum}, but as long as we understand that statements made below might sometimes have to be interpreted in terms of suitable limits, such a level of precision is, for our purposes, unnecessary.

\begin{figure}[!htbp]
\begin{center}
\centerline{\includegraphics[width=0.70\textwidth, height=0.70 \textwidth]{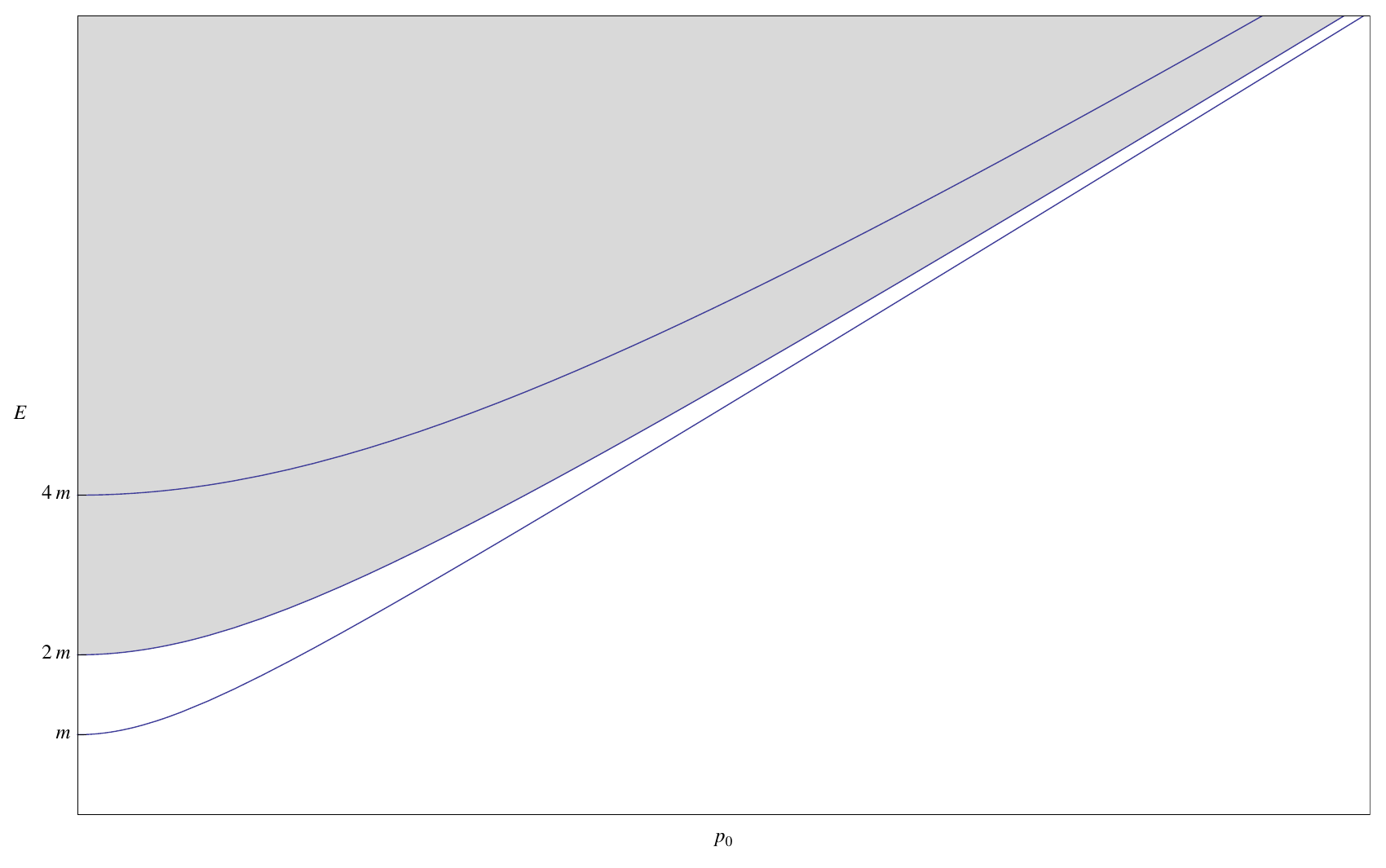}}
\caption{The kinematically accessible region for a Lorentz invariant particle of mass $m$, and a Lorentz invariant particle of mass $4m$, decaying to two identical particles of mass $m$. Note \emph{absence} of decay thresholds: The process is either allowed or forbidden in a momentum-independent manner.}
\label{F:Lorentz}
\end{center}
\end{figure}

Then the decay process (\ref{E:decay}) is kinematically allowed if and only if
\begin{equation}
E_\mathrm{min}(\p_0) \leq E_0(\p_0) \leq E_\mathrm{max}(\p_0).
\end{equation}
A decay threshold is now defined to be the edge of the kinematically allowed region. Specifically, an \emph{enabling threshold} is defined by the condition
\begin{equation}
E_\mathrm{min}(\p_0) = E_0(\p_0),
\end{equation}
and a \emph{saturation threshold} is defined by the condition
\begin{equation}
E_0(\p_0) = E_\mathrm{max}(\p_0).
\end{equation}
These thresholds are typically 2-surfaces in 3-momentum space.
\begin{itemize}
\item 
Note that $E_\mathrm{max}(\p_0)$ might trivially be infinite if any one of the energy-momentum relations does not saturate at large 3-momentum, in which case no useful upper bound, and hence no saturation threshold, would be obtained.  This is the case for instance in standard special relativity, in standard Galilean kinematics, in the Coleman--Glashow energy-momentum relation (\ref{E:Coleman}), and for $\eta>0$ in the Mattingly--Jacobson--Liberati energy-momentum relation~(\ref{E:Mattingly}.)
\item
Note that in standard special relativity the enabling threshold is also trivial --- one need merely go into the centre-of-momentum frame to see that the decay is kinematically allowed if and only if
\begin{equation}
m_0 \geq \sum_{n=1}^n m_i.
\end{equation}
\end{itemize}
That is, decay thresholds are trivial in the case of exact Lorentz invariance, (see figure~\ref{F:Lorentz}) and only become interesting if there are deviations from Lorentz invariance.

\subsection{Thresholds in momentum space} 
\label{S:momentum}

For a graphical understanding of the situation it is useful to pick some (arbitrary but fixed) direction $\hat\p$ in momentum space, write $\p_0 = p_0 \;\hat\p$, and for each direction $\hat \p$ consider the three curves:
\begin{eqnarray}
\C_+(\hat\p) &=&  \{E_\mathrm{max}(p_0 \;\hat\p), p_0\};
\\
\C_0(\hat\p) &=&  \{E_0(p_0 \;\hat\p), p_0\};
\\
\C_-(\hat\p) &=&  \{E_\mathrm{min}(p_0 \;\hat\p), p_0\}.
\end{eqnarray}
Note that if the individual energy-momentum relations are isotropic, (rotationally invariant, spherically symmetric), then these curves $\C_{+/0/-}$ will be independent of the direction $\hat\p$. If Lorentz invariance is violated, isotropy would at best occur only in the preferred (aether) frame, so in general it is safer to \emph{not} make any such assumption.

\begin{figure}[!htbp]
\begin{center}
\centerline{\includegraphics[width=1.00\textwidth, height=0.750 \textwidth]{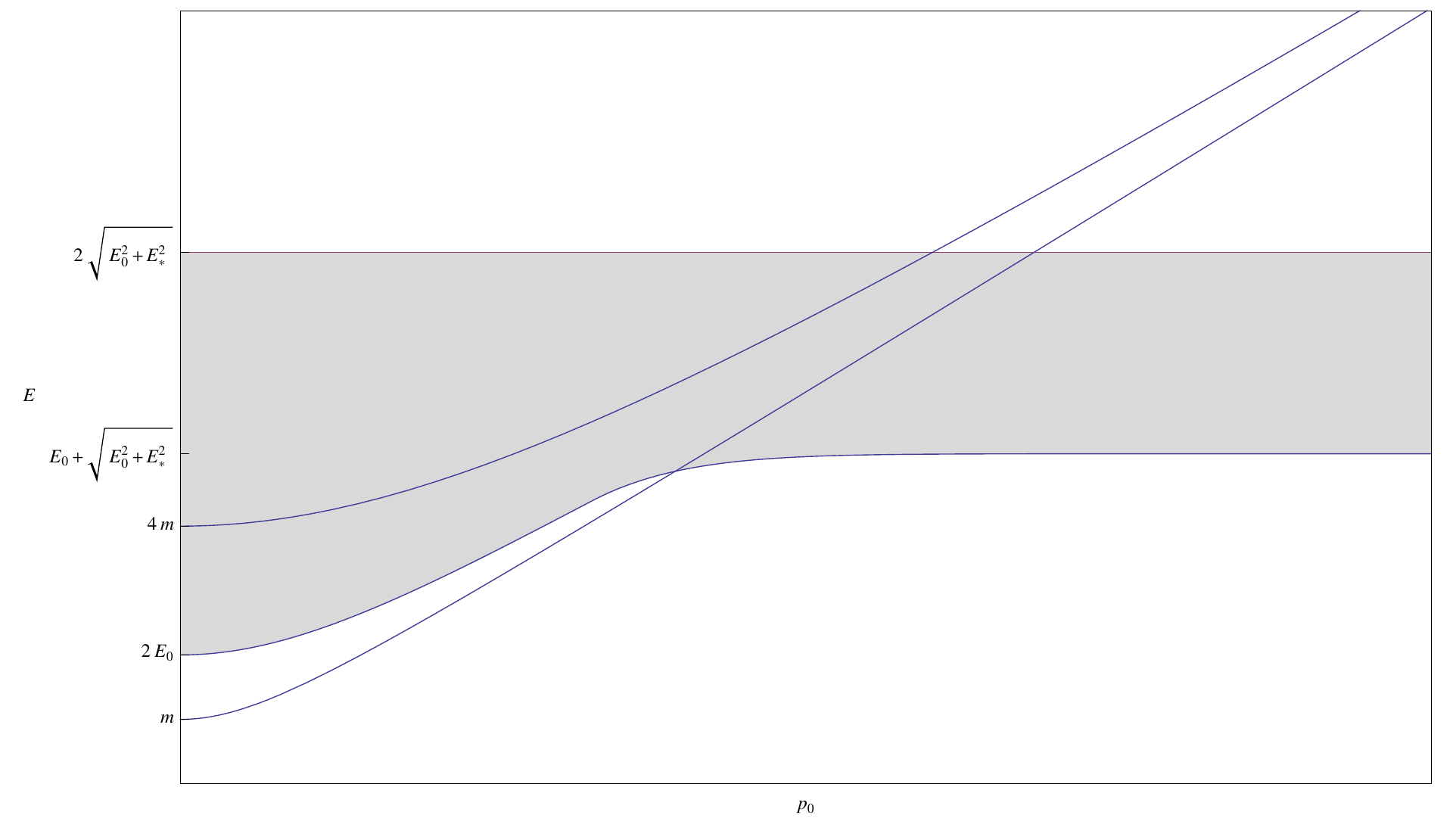}}
\caption{The kinematically accessible region for a Lorentz invariant particle of mass $m$, and a Lorentz invariant particle of mass $4m$, decaying to two identical particles with energy-momentum relation $E=\sqrt{E_0^2 + E_*^2 \tanh \left( {p^2 c^2}/{E_*^2} \right)}$. For the mass $m$ particle note the presence of both enabling and saturation thresholds. For the  mass $4m$ particle only the saturation threshold survives.}
\label{F:enabling-saturation}
\end{center}
\end{figure}

The kinematically accessible region for the decay products (assuming only conservation of 3-momentum) is the region between the curves $\C_-$ and $\C_+$. Kinematically allowed decays correspond to that portion of $\C_0$ that lies in the region between the curves $\C_-$ and $\C_+$.
Enabling thresholds occur whenever the curve $\C_0$ intersects the curve $\C_-$, saturation thresholds occur whenever the curve $\C_0$ intersects the curve $\C_+$, see figure~\ref{F:enabling-saturation}.

It is additionally useful to distinguish \emph{lower} and \emph{upper} thresholds. A lower threshold occurs when, as a function of increasing $p_0$, the curve $\C_0$ enters the kinematically accessible region, and an upper threshold occurs when, as a function of increasing $p_0$, the curve $\C_0$ leaves the kinematically accessible region.

\begin{figure}[!htbp]
\begin{center}
\centerline{\includegraphics[width=1.00\textwidth, height=0.750 \textwidth]{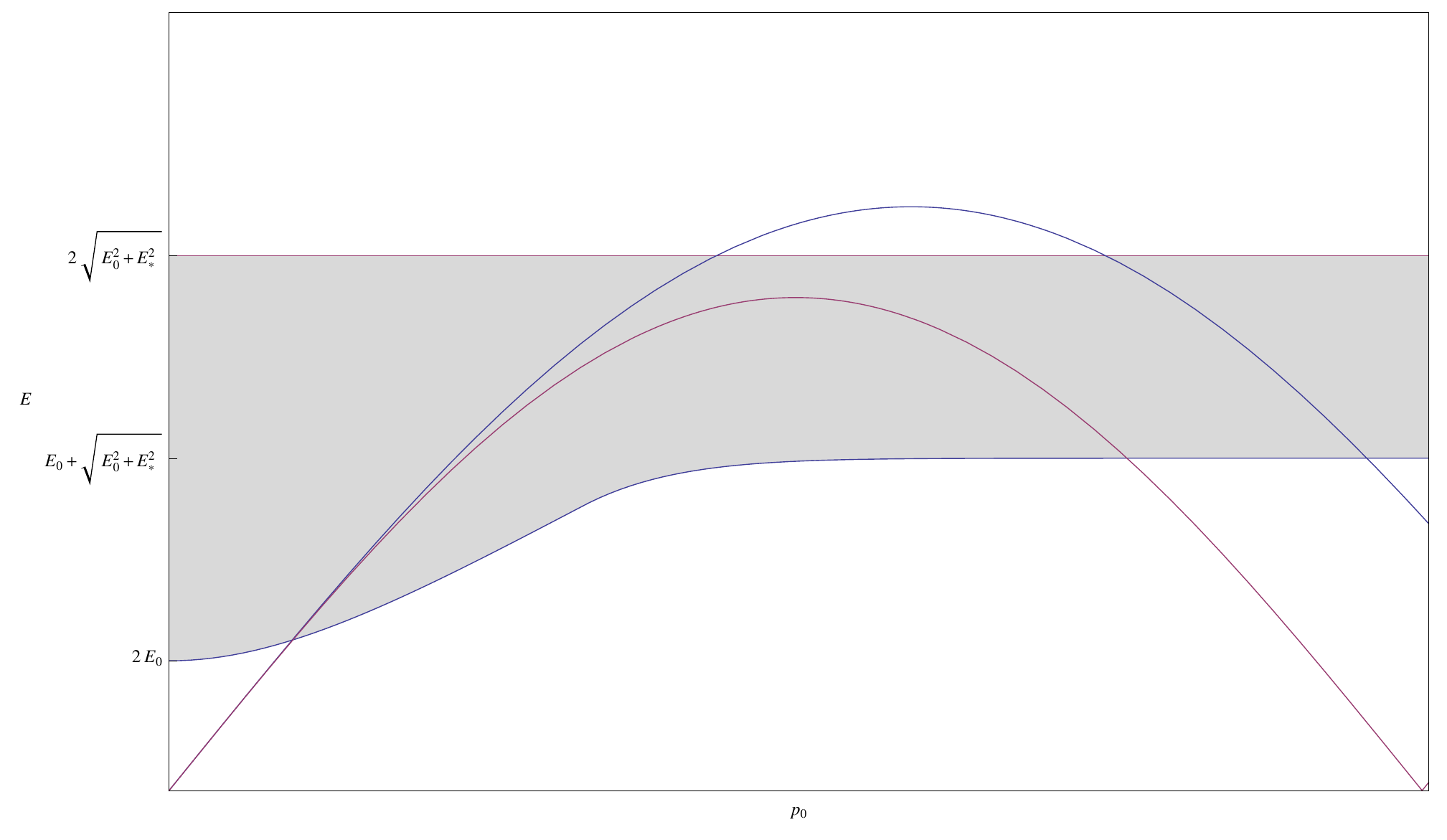}}
\caption{The kinematically accessible region for initial particles with a ``lattice-like'' energy-momentum relation $E= (\hbar c/a) |\sin(pa/\hbar)|$, (with two distinct values of the ``lattice spacing'' $a$), decaying to two identical particles with ``\emph{tanh}-like'' energy-momentum relation $E=\sqrt{E_0^2 + E_*^2 \tanh \left( {p^2 c^2}/{E_*^2} \right)}$. For small lattice spacing note the presence of four thresholds: In order they are lower enabling, upper saturation, lower saturation, and upper enabling thresholds. For larger lattice spacing only the lower enabling and upper enabling thresholds survive.}
\label{F:lower-upper}
\end{center}
\end{figure}

\begin{figure}[!htbp]
\begin{center}
\centerline{\includegraphics[width=0.70\textwidth, height=0.70 \textwidth]{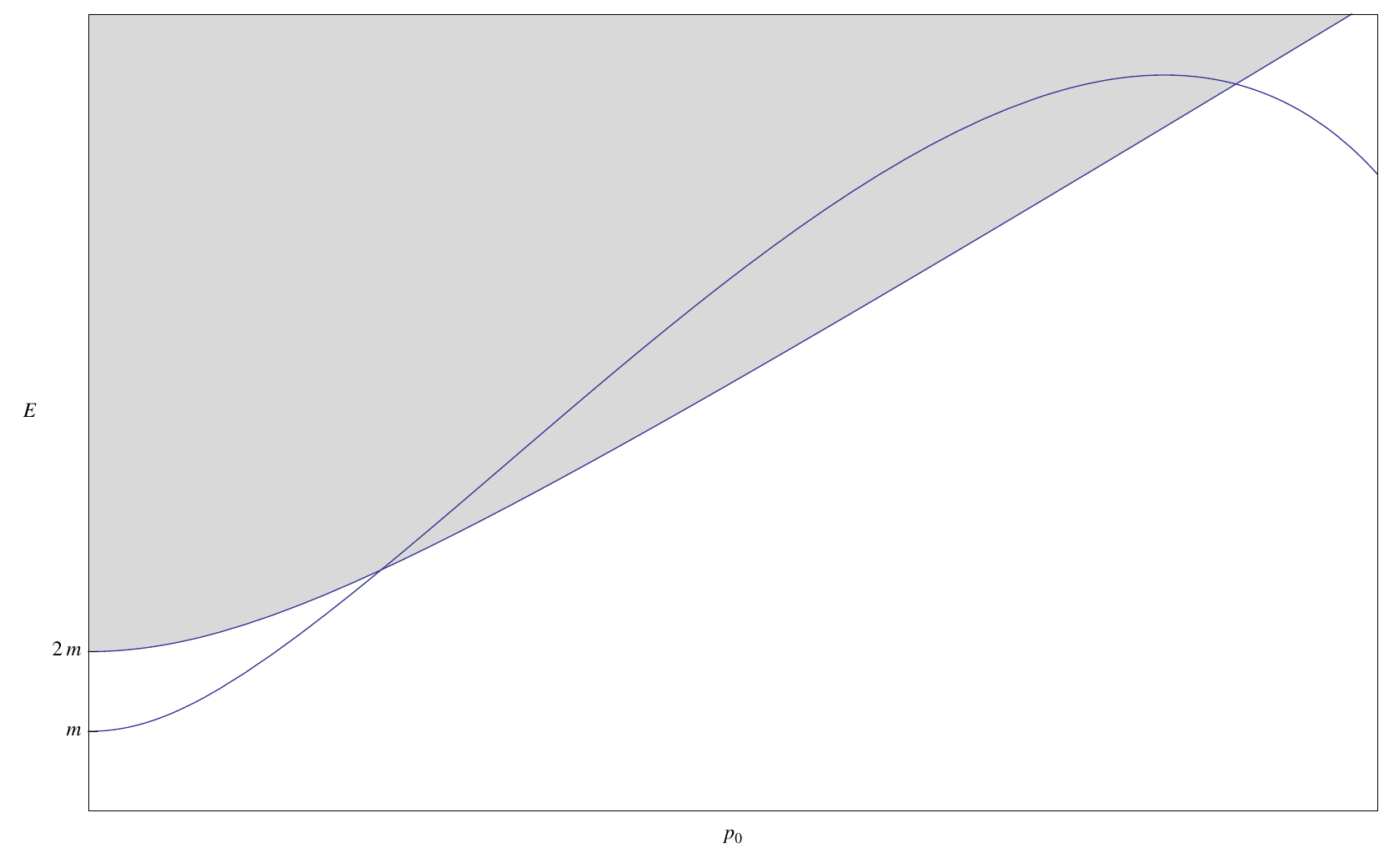}}
\caption{The kinematically accessible region for a particle with a mass $m$ and a polynomial energy-momentum relation, decaying to two identical Lorentz invariant particles of mass $m$. Note presence of both lower and upper enabling thresholds, but no saturation thresholds. 
}
\label{F:Lorentz2}
\end{center}
\end{figure}

\begin{figure}[!htbp]
\begin{center}
\centerline{\includegraphics[width=0.70\textwidth, height=0.70 \textwidth]{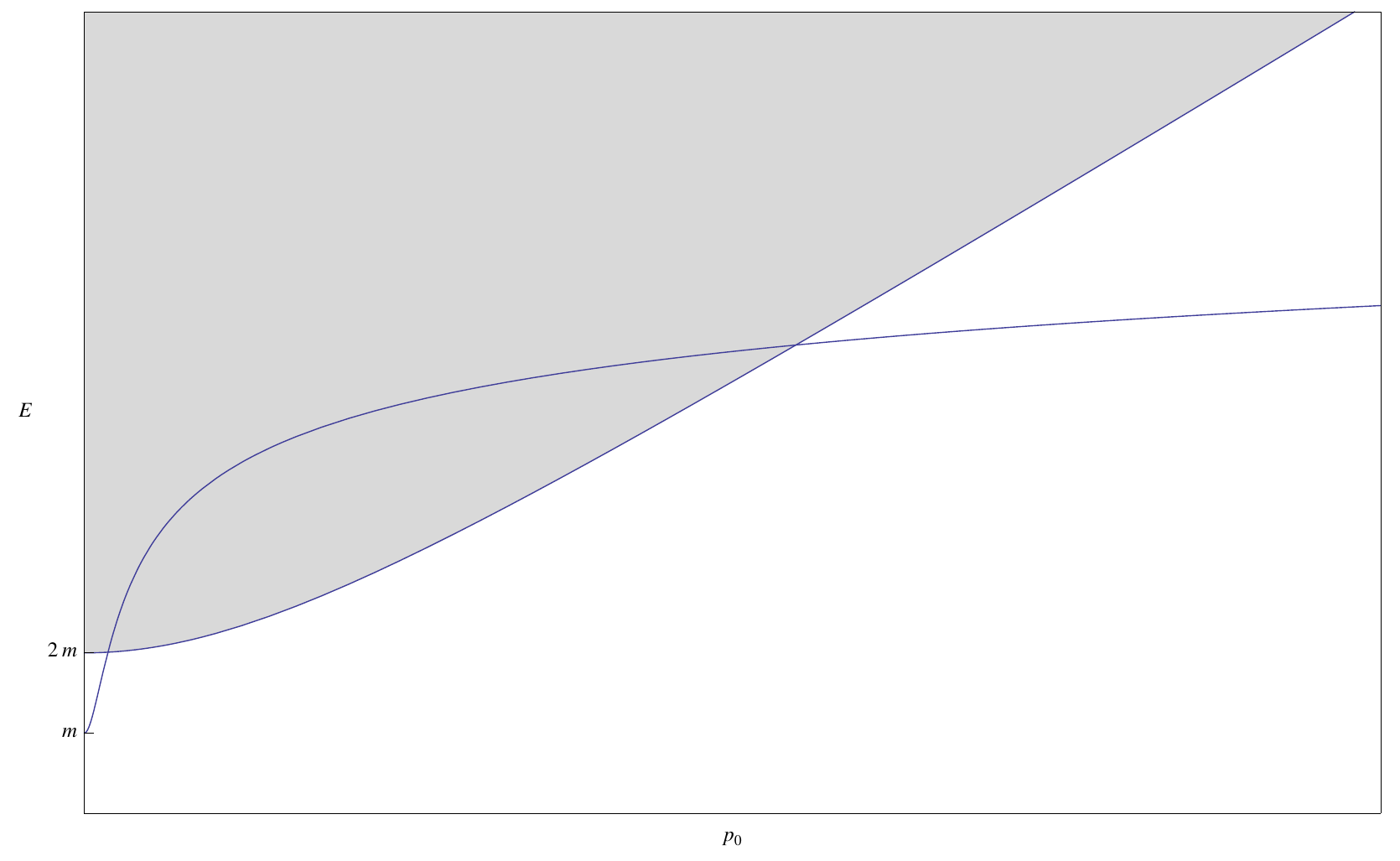}}
\caption{The kinematically accessible region for a particle of mass $m$ with a complicated but monotonic energy-momentum relation, decaying to two identical Lorentz invariant particles of mass $m$. Note presence of both lower and upper enabling thresholds, but no saturation thresholds. 
}
\label{F:Lorentz3}
\end{center}
\end{figure}

There are some quite general results for the curves $\C_\pm$ that are not too difficult to establish. Consider for simplicity a 2-particle final state, or a 2-particle subsystem of a $n$-body final state. Then consider the (restricted) set of curves
\begin{equation}
\C_\epsilon(\hat\p) = \Big\{ E_1(\epsilon p_0 \hat\p) + E_2([1-\epsilon] p_0 \hat\p), p_0 \Big\} \qquad \forall\epsilon\in\R,
\end{equation}
and (now with $\hat\p\cdot\hat\p_\perp=0$) the more general set of curves
\begin{equation}
\C_{\epsilon,\delta}(\hat\p,\hat\p_\perp) = \Big\{ E_1(\epsilon p_0 \hat\p+ \delta \, \hat\p_\perp) 
+ E_2([1-\epsilon] p_0 \hat\p - \delta \, \hat\p_\perp), p_0 \Big\} \qquad \forall\epsilon, \delta \in\R.
\end{equation}
\emph{All} of these curves lie between $\C_-$ and $\C_+$, and can be used to quickly sketch out the kinematically allowed region.   In particular the curve
\begin{equation}
\Big\{ E_1(p_0 \hat\p/2) + E_2(p_0 \hat\p/2), p_0 \Big\}
\end{equation}
corresponds to sharing momentum equally between the two particles,  and so automatically lies between $\C_-$ and $\C_+$. Perhaps less obviously the two curves
\begin{equation}
\Big\{ E_1({p_0 \hat\p}) + E_2(\0), p_0 \Big\}
\qquad \hbox{and} \qquad 
\Big\{ E_1(\0) + E_2({p_0 \hat\p}), p_0 \Big\}
\end{equation}
correspond to putting all available 3-momentum into particle 1 or particle 2 respectively, and both these curves automatically lie between $\C_-$ and $\C_+$. By considering the limits $\epsilon \to\pm\infty$ we also see that the two \emph{horizontal} lines
\begin{equation}
\Big\{ E_1(\pm\infty\hat\p) + E_2(\mp\infty\hat\p), p_0 \Big\}
\end{equation}
lie in the kinematically allowed region. (These last two curves are most useful when the energy-momentum relations saturate at large 3-momentum.) Finally note that the \emph{vertical} line 
\begin{equation}
\Big\{ E_1(p_0 \hat\p) + E_2(-p_0 \hat\p), 0 \Big\}
\end{equation}
also lies entirely within the kinematically allowed region.  These observations allow one to quickly sketch key features of the kinematically allowed region. Some graphical experiments will quickly convince one that in general the kinematically allowed region need not be convex, nor need the curves $\C_\pm$ necessarily be monotonic. 
If one is willing to make more specific assumptions concerning the energy-momentum relations, only then can much more be said about $\C_\pm$ and $E_\mathrm{max/min}(\p_0)$. For instance:
\begin{itemize}

\item
For the lattice energy-momentum relation of equation (\ref{E:lattice}) we have $E_i(\p_i) \leq \hbar c/a$, so provided all final state decay products see the \emph{same} lattice, we have
$E_\mathrm{max}(\p_0) \leq {n\hbar c/a} $, and so the curve $\C_+$ will be nontrivial.

\item
For the \emph{tanh}-type energy-momentum relation of equation (\ref{E:tanh}) it is easy to check that
\begin{equation}
E_\mathrm{max}(\p_0) = \sum_{i=1}^n \sqrt{E_{0,i}^2 + E_{*,i}^2}.
\end{equation}
Then $\C_+$ is a simple horizontal line. $\C_-$ is however quite nontrivial, see figure \ref{F:enabling-saturation}.
\end{itemize}
Once one abandons isotropy --- in particular azimuthal isotropy around the chosen direction $\hat\p$ in momentum space --- then a fuller analysis using the curves $\C_{\epsilon,\delta}(\hat\p,\hat\p_\perp)$ will be necessary.  Formally
\begin{equation}
\C_-(\hat \p) =  \min_{\epsilon\in \R}  \min_{\hat\p_\perp} \min_{\delta\in\R} \C_{\epsilon,\delta}(\hat\p,\hat\p_\perp),
\end{equation}
and
\begin{equation}
\C_+(\hat \p) =  \max_{\epsilon\in \R}  \max_{\hat\p_\perp} \max_{\delta\in\R} \C_{\epsilon,\delta}(\hat\p,\hat\p_\perp).
\end{equation}
Typically the curves $\C_\pm(\hat\p)$  will piecewise consist of segments of some specific curves chosen from the $\C_{\epsilon,\delta}(\hat\p,\hat\p_\perp)$. 
The key message to extract from the discussion is this: 
Once exact Lorentz invariance is lost the kinematically allowed region can become extremely complicated.

\subsection{Lagrange multiplier techniques}
\label{S:lagrange}

Motivated by the more restricted analysis of Coleman and Glashow~\cite{Coleman:1998ti}, which we now significantly generalize, further  technical progress can best be made by introducing Lagrange multipliers and considering extrema (at fixed $\p_0$) of the function
\begin{equation}
\E(\p_0;\p_i,\blambda) = \sum_{i=1}^n E_i(\p_i) - \blambda\cdot \left( \sum_{i=1}^n \p_i - \p_0\right).
\end{equation}
All minima used to determine $E_\mathrm{min}(\p_0)$, or maxima used to determine $E_\mathrm{max}(\p_0)$, will be extrema of the function $\E(\p_0;\p_i,\blambda)$ (though not necessarily vice versa). So extrema of $\E(\p_0;\p_i,\blambda)$ will provide information concerning thresholds. 

But all extrema of $\E$ satisfy
\begin{equation}
{\partial E_i \over \partial \p_i }=\blambda.
\end{equation}
In view of the specific Hamilton equation $\vv = \dot\x = \partial E/\partial\p$, this
implies that at any extremum $\vv_i = \blambda = \vv_\mathrm{out}$ --- all output velocities for the decay products are equal at any extremum, so in particular all output velocities are equal for the specific configuration of decay product 3-momenta $\p_i$ that define $E_\mathrm{min}(\p_0)$ and $E_\mathrm{max}(\p_0)$. 
That is, we have the very general result that at threshold all final state particles move with the same 3-velocity.
Furthermore for any extremum we also have
\begin{equation}
{\partial \E(\p_0; \p_i, \blambda) \over\partial\p_0} = \blambda = \vv_\mathrm{out},
\end{equation}
so in particular this will also be true for $E_\mathrm{min}(\p_0)$ and $E_\mathrm{max}(\p_0)$.  That is: For the specific configuration of decay product 3-momenta $\p_i$ that define $E_\mathrm{min}(\p_0)$ and $E_\mathrm{max}(\p_0)$ we have
\begin{eqnarray}
{\partial E_\mathrm{min}\over \partial \p_0} =  \vv_\mathrm{out, min} = {\partial E_i \over \partial \p_i },
\end{eqnarray}
and
\begin{eqnarray}
{\partial E_\mathrm{max}\over \partial \p_0} =  \vv_\mathrm{out, max} = {\partial E_i \over \partial \p_i },
\end{eqnarray}
respectively.
Physically this implies that at any threshold (regardless of whether it is an enabling threshold or a saturation threshold, or a lower or upper threshold) all decay products will be moving at the same physical 3-velocity.  This does \emph{not} necessarily imply that the 3-momenta be related in any simple way, in general the 3-momenta need not even be collinear. At threshold we can define the incoming 3-velocity as
\begin{eqnarray}
{\partial E_0\over \partial \p_0} =  \vv_\mathrm{in},
\end{eqnarray}
but with the techniques currently at hand there is in general no simple relation between $\vv_\mathrm{in}$ and $ \vv_\mathrm{out}$. The best we can currently do is this:
If we look along a particular direction $\hat\p$ in 3-momentum space (with $\p_0=p_0\;\hat\p$) then:
\begin{itemize}
\item 
At a \emph{lower enabling} threshold $ \hat\p\cdot \vv_\mathrm{in}(\p_0) \geq \hat\p\cdot \vv_\mathrm{out}(\p_0)$.
\item 
At an \emph{upper enabling} threshold $ \hat\p\cdot \vv_\mathrm{in}(\p_0) \leq \hat\p\cdot \vv_\mathrm{out}(\p_0)$.
\item 
At a \emph{lower saturation} threshold $ \hat\p\cdot \vv_\mathrm{in}(\p_0) \leq \hat\p\cdot \vv_\mathrm{out}(\p_0)$.
\item 
At an \emph{upper saturation} threshold $ \hat\p\cdot \vv_\mathrm{in}(\p_0) \geq \hat\p\cdot \vv_\mathrm{out}(\p_0)$.
\item
If  $ \hat\p\cdot \vv_\mathrm{in}(\p_0) = \hat\p\cdot \vv_\mathrm{out}(\p_0)$ then the curve $\C_0(\hat\p)$ touches the kinematically allowed region tangentially. One should look at higher derivatives to determine the nature of the threshold.  If  the curve $\C_0(\hat\p)$ touches the kinematically allowed region only at an isolated point, then we would hesitate to call this any kind of threshold. (These isolated points could nevertheless be interesting in their own right.)
\end{itemize}
In the case of an isotropic energy-momentum relation this discussion simplifies. The momentum $\p$ is then parallel (or at worst anti-parallel) to the velocity $\vv$ and so: 
\begin{itemize}
\item 
At a \emph{lower enabling} threshold $ v_\mathrm{in}(\p_0) \geq  v_\mathrm{out}(\p_0)$.
\item 
At an \emph{upper enabling} threshold $v_\mathrm{in}(\p_0) \leq  v_\mathrm{out}(\p_0)$.
\item 
At a \emph{lower saturation} threshold $ v_\mathrm{in}(\p_0) \leq  v_\mathrm{out}(\p_0)$.
\item 
At an \emph{upper saturation} threshold $v_\mathrm{in}(\p_0) \geq  v_\mathrm{out}(\p_0)$.
\item 
The special case  $ v_\mathrm{in}(\p_0) =  v_\mathrm{out}(\p_0)$ should be analyzed carefully by looking at higher derivatives. 
This  might correspond to an ``isolated point at which the decay is allowed''; we would then hesitate to call this any kind of threshold.
\end{itemize}

\subsection{Thresholds in terms of energy} 
\label{S:energy}

With some additional technical machinery we can rephrase the decay thresholds in terms of energy rather than 3-momentum.  Some aspects of the analysis are more complicated, but we will now be able to deduce (at threshold) that $\vv_\mathrm{in}$ and $ \vv_\mathrm{out}$ are parallel/anti-parallel.
Let us now define
\begin{equation}
\P_\mathrm{in}(E_0) = \Big\{ \p_0 : E_0(\p_0)= E_0\Big\},
\label{E:P_in}
\end{equation}
which is the set of all possible total 3-momenta given the input energy $E_0$. 
Now consider
\begin{equation}
\P_\mathrm{out}(\P_\mathrm{in}(E_0)).
\end{equation}
This is the set of planes (affine subspaces) of co-dimension 3 in $\R^{3n}$ consisting of all possible output 3-momenta compatible with the specified input energy $E_0$.  Then 
\begin{equation}
E_\mathrm{out}(\P_\mathrm{out}(\P_\mathrm{in}(E_0)))
\end{equation}
is the set of all possible 3-momentum-conserving output energies for input energy $E_0$. This will be a connected interval in $\R$. The decay process is then kinematically allowed if and only if
\begin{equation}
E_0 \in E_\mathrm{out}(\P_\mathrm{out}(\P_\mathrm{in}(E_0))).
\end{equation}
That is
\begin{equation}
E_\mathrm{min}(\P_\mathrm{out}(\P_\mathrm{in}(E_0))) \leq E_0 \leq E_\mathrm{max}(\P_\mathrm{out}(\P_\mathrm{in}(E_0))),
\end{equation}
where by this we mean
\begin{equation}
\min_{\p_0\in \P_\mathrm{in}(E_0)} E_\mathrm{min}(\p_0) \leq E_0 \leq \max_{\p_0\in \P_\mathrm{in}(E_0)} E_\mathrm{max}(\p_0).
\end{equation}
Notice now that we are also extremizing over the 3-momenta $\p_0$ compatible with the fixed initial energy $E_0$. 
Enabling thresholds will then occur at
\begin{equation}
\min_{\p_0\in \P_\mathrm{in}(E_0)} E_\mathrm{min}(\p_0) = E_0,
\end{equation}
and saturation thresholds at
\begin{equation}
 E_0 = \max_{\p_0\in \P_\mathrm{in}(E_0)} E_\mathrm{max}(\p_0).
\end{equation}
Either one of these thresholds can be characterized in terms of extrema of the related function
\begin{equation}
\E(E_0;\p_i,\p_0,\blambda,\zeta)
= \sum_{i=1}^n E_i(\p_i) - \blambda \cdot \left[\sum_{i=1}^n \p_i - \p_0\right] 
- \zeta\left[ E_0(\p_0) - E_0\right],
\end{equation}
where we now introduce two Lagrange multipliers, $\blambda$ and $\zeta$, and we extremize over $(\p_i,\p_0,\blambda,\zeta)$ while keeping $E_0$ fixed. Extremality with respect to the $\p_i$ yields
\begin{equation}
\vv_i = \blambda = \vv_\mathrm{out},
\end{equation}
whereas extremality with respect to $\p_0$ yields
\begin{equation}
\zeta\;  \vv_\mathrm{in} = \blambda. 
\end{equation}
Since the \emph{sign} and magnitude of $\zeta$ is unconstrained, this implies that at threshold the input velocity  $\vv_\mathrm{in}$ is  either parallel or anti-parallel to the common $\vv_\mathrm{out}$ of all the output particles:
\begin{equation}
\zeta \;  \vv_\mathrm{in} = \vv_\mathrm{out}.
\end{equation}
Note that we have gotten at least this far without assuming either spherical symmetry or any form of monotonicity.

\subsection{Asymmetric thresholds}
\label{S:asymmetric}

A particularly peculiar feature of Lorentz violating thresholds is the potential occurrence of asymmetric thresholds, where two identical decay particles might at threshold have unequal 3-momenta while traveling at the same 3-velocity. (This phenomena was noted, in a more limited context, in reference~\cite{Mattingly:2002ba}.) 
The point is that while $\dot\x(\p)$ is by assumption well defined, the inverse function $\p(\dot\x)$ may be multi-valued. If this happens at threshold then two identical particles in the decay channel will have equal velocities but unequal momenta.
Of course such multi-valued behaviour implies a multi-valued Lagrangian $L(\dot\x) = \p(\dot\x) \cdot \dot\x - E(\p(\dot\x))$, which one may wish to exclude from any fundamental theory on physical grounds.   (Such behaviour in an effective field theory is not particularly problematic.) 
To characterize when this can and cannot happen, note that \emph{local} invertability of $\dot\x(\p)$ requires the Jacobian matrix
\begin{equation}
{\partial \dot\x\over\partial\p}
\end{equation}
to be nonsingular. Equivalently the Hessian matrix
\begin{equation}
{\partial^2 E \over\partial\p\, \partial\p}
\end{equation}
should be nonsingular. 

\emph{Global} invertability of $\dot\x(\p)$ requires global non-singularity of the Hessian matrix. If we now add the extremely mild constraint that the Hessian matrix be positive definite at zero momentum (which is required to have any sensible Newtonian or Lorentzian limit at low momentum) then global invertability of  $\dot\x(\p)$ requires the Hessian matrix to be globally positive definite. 
But a globally positive definite Hessian matrix implies convexity of the energy-momentum relation $E(\p)$. 

Thus the existence (or not) of asymmetric thresholds is ultimately related to failures (or not) of the  convexity of the energy-momentum relation $E(\p)$.  (For isotropic energy-momentum relations, this condition was phrased in terms of a positive curvature condition in reference~\cite{Mattingly:2002ba}.) 
This is why we can never get asymmetric thresholds in standard (non-tachyonic Lorentz invariant) special relativity, and why we do run the risk of asymmetric thresholds with (for example) lattice-type, \emph{tanh}-type, and polynomial or rational polynomial energy-momentum relations.  Thus the asymmetric threshold phenomena encountered by Mattingly, Jacobson, and Liberati in reference~\cite{Mattingly:2002ba} is seen to have much wider applicability than the situations they considered.

\subsection{Some examples}
\label{S:examples}

\begin{figure}[!htbp]
\begin{center}
\centerline{\includegraphics[width=0.50\textwidth, height=0.50 \textwidth]{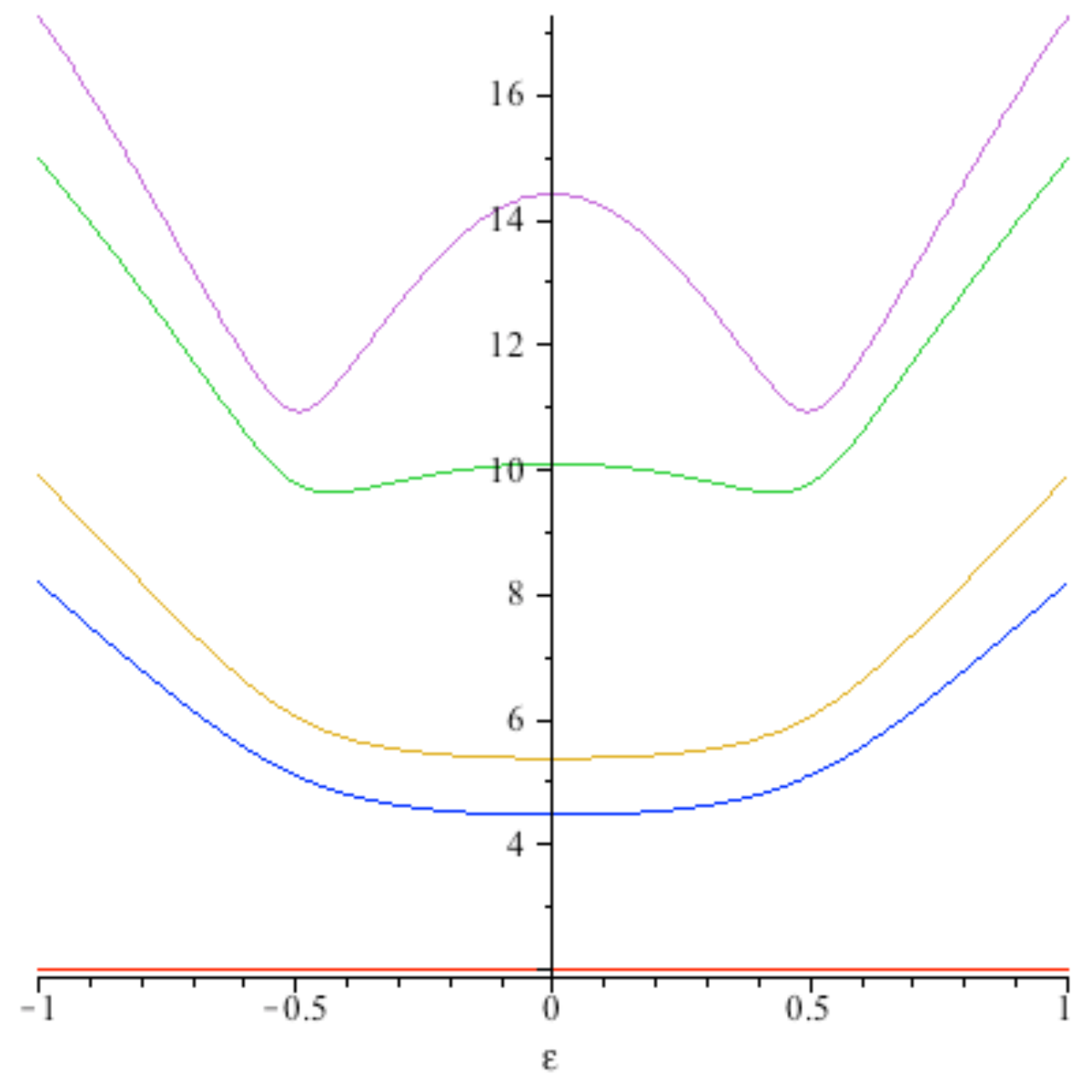}}
\caption{Energy as a function of $\epsilon$ for equation~(\ref{E:tanh2}) holding $E_0$ and $E_*$ fixed and for five distinct values of $p_0$. Note how the location of the minimum (and hence the threshold) shifts from $\epsilon=0$ (a symmetric threshold) to $\epsilon \approx \pm1/2$ (an asymmetric threshold).  When the minimum occurs at $\epsilon \approx \pm1/2$ almost all of the output momentum goes into one of the two identical particles, and almost none into the other. 
}
\label{F:asymmetric}
\end{center}
\end{figure}

As an example of what can happen with asymmetric thresholds, it is quite possible for two identical particles to be emitted with almost all the momentum going into one particle, and almost none into the second particle. For instance if one takes two decay product particles obeying the \emph{tanh}-type energy-momentum relation of (\ref{E:tanh}), and shares the input momentum $p_0$ in the fractions $({1\over2}\pm\epsilon) p_0$, then the final state energy is
\begin{equation}
E(p_0,\epsilon) =  \sqrt{E_0^2 + E_*^2 \; \tanh\left( { [{1\over2}+\epsilon]^2 p_0^2 c^2\over E_*^2} \right) } + \sqrt{E_0^2 + E_*^2 \; \tanh\left( { [{1\over2}-\epsilon]^2 p_0^2 c^2\over E_*^2} \right) }.
\label{E:tanh2}
\end{equation}
Depending on the precise ratios between $E_0$, $E_*$, and $p_0$, this can be minimized at $\epsilon=0$ or near $\epsilon=1/2$. See figure~\ref{F:asymmetric}. 

Another highly nontrivial example, based roughly on equation~(\ref{E:anisotropic}), is to take
\begin{equation}
E= \sqrt{ E_0^2 + c^2(\p_x^2+\p_y^2) + {c^2\over3p_*^4}\left\{ (\p_z^2-p_*^2)^3 - p_*^6 \right\} }. 
\end{equation}
This energy momentum relation is carefully chosen to be isotropic at low momentum, to have nice behaviour in the $x$ and $y$ directions, and to behave ``interestingly'' in the $z$ direction. Consider now a particle that moves in the $x$ direction so its original momentum is $(p_0,0,0)$. Let it now decay into two identical particles of the type discussed above. Imposing 3-momentum conservation, the energy of the final state will be minimized when the final state particles have 3-momenta $(p_0/2, 0, \pm p_*)$. So at threshold the initial state and two final state 3-momenta are pointing in three different directions. The 3-velocities of the two final state particles will however be equal, and will point along the $x$ axis. 

These examples should be viewed as illustrations of the unusual phenomena that can occur once strict Lorentz invariance is violated.

\subsection{Monotonicity}
\label{S:monotonicity}

What, if anything, can we say about monotonic energy-momentum relations? (This is a common but not universal simplifying assumption.) 
Consider a set of fixed directions $\hat\p_i$ and take
\begin{equation}
E_i(\p_i) = E_i(p_i \; \hat \p_i).
\end{equation}
For each individual direction $\hat\p_i$ we can define can define monotonicity in terms of the magnitude $p_i$. Monotonicity means
\begin{equation}
{d E_i(p_i \; \hat\p_i)\over d p_i} > 0,
\end{equation}
but by the chain rule this implies
\begin{equation}
{d E_i(\p_i)\over d \p_i} \cdot \hat\p_i > 0.
\end{equation}
That is
\begin{equation}
\vv_i(\p_i) \cdot \p_i > 0.
\end{equation}
But then, by our previous arguments
\begin{equation}
{d E_\mathrm{min/max}\over d \p_0} \cdot \p_0 
= \vv_\mathrm{out}  \cdot \p_0 
= \vv_\mathrm{out}  \cdot \left(\sum_{i=1}^n \p_i\right) 
= \sum_{i=1}^n \vv_\mathrm{out}  \cdot \p_i 
= \sum_i \vv_i  \cdot \p_i > 0.
\end{equation}
That is, as long as the individual $E_i(p_i \; \hat \p_i)$ are monotonic functions of the $p_i$, then $E_\mathrm{min/max}(p_0 \; \hat\p_0)$ is also monotonic as a function of $p_0$, and so the curves $\C_\pm$ bounding the kinematically allowed region will be monotonic. (Monotonicity of the boundary curves $\C_\pm$ can fail, and quite often will fail, if even one of the final state particles has a non-monotonic energy-momentum relation.) 

\subsection{Isotropy}
\label{S:isotropy}

If all the energy-momentum relations are isotropic (in the preferred (aether) frame) then 
\begin{equation}
\vv_i \propto \p_i; \qquad \hbox{and} \qquad \vv_{\mathrm{in}} \propto \p_0.
\end{equation}
So all the 3-momenta $\p_i$ and $\p_0$ are either parallel or anti-parallel to their corresponding 3-velocities at threshold. Consequently, in view of the more general results deduced above, 
\begin{equation}
\p_i  \propto \p_0.
\end{equation}
That is, all 3-momenta are either parallel or anti-parallel to each other at threshold.

\subsection{Monotonicity plus isotropy}
\label{S:monotonicity-plus-isotropy}

Only if we assume both isotropy and monotonicity can we deduce that the proportionality constants in the previous subsection are positive. In this case all 3-momenta and 3-velocities are parallel at threshold. (This particular theorem was proven by Mattingly, Jacobson, and Liberati in~\cite{Mattingly:2002ba}.)

\section{Scattering thresholds} 
\label{S:scattering}

Much of the previous discussion of decay thresholds carries over into the discussion of scattering thresholds, but there are just enough differences to make some separate discussion worthwhile.

\subsection{2-particle collisions}
\label{S:2-particle}

Consider a 2-particle scattering process of the form 
\begin{equation}
X_A + X_B \to  X_1 + X_2 + \dots X_n
\end{equation}
involving particles of incoming momenta $\p_A$ and $\p_B$.  Set $\p_0=\p_A+\p_B$. 
We can still define both $\P_\mathrm{out}(\p_0)$ and $E_\mathrm{out}(\P_\mathrm{out}(\p_0))$, and so construct both $E_\mathrm{min}(\p_0) $ and $E_\mathrm{max}(\p_0)$. 
The scattering is kinematically allowed if and only if
\begin{equation}
E_A(\p_A) + E_B(\p_B) \in E_\mathrm{out}(\P_\mathrm{out}(\p_A+\p_B)).
\end{equation}
That is, it is kinematically allowed if and only if
\begin{equation}
E_\mathrm{min}(\p_A+\p_B) \leq E_A(\p_A) + E_B(\p_B) \leq E_\mathrm{max}(\p_A+\p_B).
\end{equation}
Thresholds occur at the boundaries of these regions, that is, at:
\begin{equation}
E_\mathrm{min}(\p_A+\p_B)= E_A(\p_A) + E_B(\p_B),
\end{equation}
and at
\begin{equation}
E_A(\p_A) + E_B(\p_B) = E_\mathrm{max}(\p_A+\p_B).
\end{equation}
To make further progress let us now define
\begin{equation}
\P_\mathrm{in}(E_A,E_B) = \Big\{ \p_0 = \p_A+\p_B : E_A(\p_A)= E_A, E_B(\p_B) = E_B\Big\},
\end{equation}
which is the set of all possible total input 3-momenta given the input energies $E_A$ and $E_B$. 
In terms of the notation (\ref{E:P_in}) developed for decay processes we can write
\begin{equation}
\P_\mathrm{in}(E_A,E_B)  = \P_\mathrm{in}(E_A)  + \P_\mathrm{in}(E_B).
\end{equation}
Now consider
\begin{equation}
\P_\mathrm{out}(\P_\mathrm{in}(E_A,E_B)).
\end{equation}
This is the set of planes (affine subspaces) of co-dimension 3 in $\R^{3n}$ consisting of all possible output 3-momenta compatible with the specified input energies $E_A$ and $E_B$.  Then 
\begin{equation}
E_\mathrm{out}(\P_\mathrm{out}(\P_\mathrm{in}(E_A,E_B)))
\end{equation}
is the set of all possible 3-momentum-conserving output energies for input energies $E_A$ and $E_B$. This will be some connected interval in $\R$. The 2-particle scattering process is then kinematically allowed if and only if
\begin{equation}
E_A+E_B \in E_\mathrm{out}(\P_\mathrm{out}(\P_\mathrm{in}(E_A,E_B))).
\end{equation}
That is
\begin{equation}
E_\mathrm{min}(\P_\mathrm{out}(\P_\mathrm{in}(E_A,E_B))) \leq E_A+E_B \leq E_\mathrm{max}(\P_\mathrm{out}(\P_\mathrm{in}(E_A,E_B))),
\end{equation}
where by this we mean
\begin{equation}
\min_{\p_0\in \P_\mathrm{in}(E_A,E_B)} E_\mathrm{min}(\p_0) \leq E_A+E_B \leq \max_{\p_0\in \P_\mathrm{in}(E_A,E_B)} E_\mathrm{max}(\p_0).
\end{equation}
Thresholds will then occur at the edges of the kinematically allowed region.

Specifically, enabling thresholds will then occur at
\begin{equation}
\min_{\p_0\in \P_\mathrm{in}(E_A,E_B)} E_\mathrm{min}(\p_0) = E_A+E_B,
\end{equation}
and saturation thresholds at
\begin{equation}
 E_A+E_B = \max_{\p_0\in \P_\mathrm{in}(E_A,E_B)} E_\mathrm{max}(\p_0).
\end{equation}
Either one of these thresholds can be characterized in terms of extrema of the related function
\begin{equation}
\E(E_A,E_B;\p_i,\p_A, \p_B,\blambda,\zeta_A,\zeta_B),
\end{equation}
where we now introduce three Lagrange multipliers, $\blambda$, $\zeta_A$, and $\zeta_B$, and set
\begin{equation}
 \E =
\sum_{i=1}^n E_i(\p_i) - \blambda \cdot \left[\sum_{i=1}^n \p_i - \p_A - \p_B\right] 
- \zeta_A\left[ E_A(\p_A) - E_A\right] - \zeta_B\left[ E_B(\p_B) - E_B\right].
\end{equation}
We now hold $(E_A,E_B)$ fixed, and extremize with respect to $(\p_i,\p_A, \p_B,\blambda,\zeta_A,\zeta_B)$.

Extremality with respect to the $\p_i$ yields
\begin{equation}
\vv_i = \blambda = \vv_\mathrm{out},
\end{equation}
whereas extremality with respect to $\p_A$ and $\p_B$ yields
\begin{equation}
\zeta_A \;  \vv_A = \blambda = \zeta_B\; \vv_B. 
\end{equation}
Since the \emph{signs} (and magnitudes) of $\zeta_A$ and $\zeta_B$ are unconstrained, this implies that at threshold the two input velocities  $\vv_A$ and $\vv_B$ are either parallel or anti-parallel, both to each other and to the common $\vv_\mathrm{out}$ of all the output particles:
\begin{equation}
\zeta_A \;  \vv_A = -\vv_\mathrm{out} = \zeta_B \; \vv_B.
\end{equation}
Note that we have gotten at least this far --- parallel/anti-parallel input velocities --- without assuming either spherical symmetry or any form of monotonicity. Without additional assumptions we can go no further.

\subsection{Incoming 3-velocities}
\label{S:incoming}

In reference~\cite{Mattingly:2002ba} Mattingly, Jacobson, and Liberati argue that assuming spherical symmetry and monotonicity of the energy-momentum relations the incoming 3-velocities must actually be anti-parallel at (enabling) threshold.  Note that in the absence of exact Lorentz invariance spherical symmetry can at best only be expected to hold in the preferred (aether) fame. 

Certainly if we assume spherical symmetry and monotonicity this result is now simple:
Assuming spherical symmetry of the individual energy-momentum relations $E_i(\p_i) = E_i(p_i)$ for the decay products, we have $E_\mathrm{min}(\p_0)= E_\mathrm{min}(p_0)$. Furthermore assuming monotonicity of the individual $E_i(p_i \; \hat\p_i)$ we have already seen that this implies monotonicity of the $E_\mathrm{min}(p_0\;\hat\p)$. Then for enabling thresholds 
the quantity
\begin{equation}
\min_{\p_0\in \P_\mathrm{in}(E_A,E_B)} E_\mathrm{min}(\p_0)
\end{equation}
is minimized when $||\p_0||$ is minimized. Now assuming spherical symmetry for the incoming particles, so that $E_{A/B}(\p_{A/B}) = E_{A/B}(p_{A/B})$, this occurs when when $\p_A$ and $\p_B$ are anti-parallel. Assuming spherical symmetry further implies $E(\p) = f({1\over2} p^2)$ so $\vv = dE/d \p = f'({1\over2} p^2) \p$. That is, $\vv$ and $\p$ are either parallel or anti-parallel. But monotonicity in turn implies $f'({1\over2} p^2) >0$.  That is, the individual $\vv$ and $\p$ are parallel.
This, in turn, implies the two incoming 3-velocities $\vv_A$ and $\vv_B$ are anti-parallel at enabling thresholds.

In contrast, for saturation thresholds we need to consider
\begin{equation}
\max_{\p_0\in \P_\mathrm{in}(E_A,E_B)} E_\mathrm{max}(\p_0)
\end{equation}
Assuming spherical symmetry and monotonicity for the decay products this quantity is now maximized when $||\p_0||$ is maximized.  But following the argument above, by assuming spherical symmetry and monotonicity for the input particles, this in turn implies that the incoming 3-momenta $\p_A$ and $\p_B$ are parallel at saturation thresholds, which in turn implies that the incoming 3-velocities $\vv_A$ and $\vv_B$ are parallel at saturation thresholds.

Both spherical symmetry and monotonicity are essential to these results. That is: Merely deducing that the incoming velocities are parallel/antiparallel at threshold is a generic result common to all Hamiltonian-based particle kinematics in a (Lorentz violating) homogeneous spacetime.  To go further and assert that incoming 3-velocities are anti-parallel at enabling thresholds, and parallel at saturation thresholds, requires the very much stronger assumptions of spherical symmetry and monotonicity.  Our results are briefly summarized in tables \ref{T:1}--\ref{T:3}.

\begin{table}[!htbp]
\medskip
\begin{center}
{\bf Summary of threshold behaviour: Final state particles }

\medskip
\begin{tabular}{|c|c|c|c|}
\hline
\hline
 &  generic & isotropic & isotropic+monotonic \\
 \hline
 \hline
3-velocities & equal & equal &equal\\
3-momenta & uncorrelated & collinear & parallel\\
\hline
\hline
 \end{tabular}
 \end{center}
\caption{\label{T:1}Behaviour at threshold for the 3-velocities and 3-momenta of outgoing final-state particles (compared to each other) under various assumptions.}

\end{table}

\begin{table}[!htbp]
\medskip
\begin{center}
{\bf Summary of threshold behaviour: Decay --- Initial state particle. }

\medskip
\begin{tabular}{|c|c|c|c|}
\hline
\hline
 &  generic & isotropic & isotropic+monotonic \\
\hline
\hline
3-velocity & collinear & collinear & collinear\\
3-momentum & uncorrelated & collinear & parallel\\

\hline
\hline
 \end{tabular}
 \end{center}
\caption{\label{T:2}Behaviour at threshold for the 3-velocity and 3-momentum of the initial decaying particle (as compared to the final state decay product particles)  under various assumptions.}

\end{table}

\begin{table}[!htbp]
\medskip
\begin{center}
{\bf Summary of threshold behaviour: Scattering --- Initial state particles. }

\medskip
\begin{tabular}{|c|c|c|c|}
\hline
\hline
 &  generic & isotropic & isotropic+monotonic \\
\hline
\hline
3-velocities (enabling) & collinear & collinear & anti-parallel\\
3-momenta (enabling) & uncorrelated & collinear & anti-parallel\\
\hline
\hline
3-velocities (saturation) & collinear & collinear & parallel\\
3-momenta (saturation) & uncorrelated & collinear & parallel\\
\hline
\hline

 \end{tabular}
 \end{center}
\caption{\label{T:3}Behaviour at threshold for the 3-velocities and 3-momenta of the two initial state (incoming) particles (as compared to each other)  under various assumptions.}

\end{table}


\section{Conclusions} 
\label{S:conclusions}

As we have seen, abandoning Lorentz invariance carries a very high price. The kinematically allowed region, and consequent threshold structure coming from the boundaries of the kinematically allowed region, for both decay processes and 2-particle (elastic or inelastic) scattering, is \emph{much} more complicated than in the Lorentz invariant case. There are some limited number of truly general statements that one can make, but most of one's intuition has to be reassessed on a case by case basis. We have tried to carefully delineate exactly which assumptions are central to which results, concentrating on those results that depend only on the existence of a homogeneous spacetime, and adding extra assumptions only when essential to obtaining specific specialized results. 

The resulting framework is useful both in (an extremely wide class of) Lorentz violating extensions of the standard model of particle physics, and is also potentially of interest in quasi-particle settings where violations of Lorentz invariance (and even rotational invariance) are the norm rather than the exception.

Note that in the spirit of classical particle physics we have taken the Hamiltonian framework as being more fundamental, and the Lagrangian framework as derivative. As a side effect, the Lagrangians that typically arise for non-Lorentz-covariant free-particle Hamiltonians are often quite messy and unnatural. If one adopts the view that it is the Lagrangian framework that should be viewed as being more fundamental, this suggests that modified energy-momentum relations should be most naturally interpreted as effective phenomena rather than as fundamental physics.  It is ultimately for this reason that one commonly focusses on perturbative deviations from Lorentz symmetry. A central theme of this article is that such a simplification is not always the most useful thing to do, and that there is merit to analyzing thresholds in as general a setting as possible. While the analysis of this article has no direct bearing on the OPERA--MINOS observations, any serious attempt at phenomenological analysis of those observations will need to adopt a theoretical framework along the lines we have presented above.

\section*{Acknowledgments} 

This research was supported by the Marsden Fund, 
administered by the Royal Society of New Zealand.  Additionally, Valentina Baccetti acknowledges support by a Victoria University PhD scholarship, and Kyle Tate acknowledges support by a Victoria University MSc scholarship.
The authors especially wish to thank David Mattingly, Stefano Liberati, and Ted Jacobson for their comments and feedback.



\begin{thebibliography}{99}
\bibitem{Opera:2011}
  T.~Adam {\it et al.} [ OPERA Collaboration ],
  ``Measurement of the neutrino velocity with the OPERA detector in the CNGS beam'',
   [arXiv:1109.4897 [hep-ex]].
\bibitem{Minos:2007}
  P.~Adamson {\it et al.} [ MINOS Collaboration ],
  ``Measurement of neutrino velocity with the MINOS detectors and NuMI neutrino beam'',
  Phys.\ Rev.\  {\bf D76 } (2007)  072005.
  [arXiv:0706.0437 [hep-ex]].
\bibitem{AmelinoCamelia:2011dx}
  G.~Amelino-Camelia, G.~Gubitosi, N.~Loret, F.~Mercati, G.~Rosati, P.~Lipari,
  ``OPERA --- Reassessing data on the energy dependence of the speed of neutrinos'',
  [arXiv:1109.5172 [hep-ph]].
\bibitem{Giudice:2011mm}
  G.~F.~Giudice, S.~Sibiryakov, A.~Strumia,
  ``Interpreting OPERA results on superluminal neutrino'',
  [arXiv:1109.5682 [hep-ph]].
\bibitem{Cohen:2011hx}
  A.~G.~Cohen and S.~L.~Glashow,
  ``Pair Creation Constrains Superluminal Neutrino Propagation,''
  Phys.\ Rev.\ Lett.\ \ {\bf 107} (2011) 181803
  [arXiv:1109.6562 [hep-ph]].
\bibitem{Dvali:2011mn}
  G.~Dvali, A.~Vikman,
  ``Price for Environmental Neutrino-Superluminality'',
  [arXiv:1109.5685 [hep-ph]].
\bibitem{Alexandre:2011bu}
  J.~Alexandre, J.~Ellis, N.~E.~Mavromatos,
  ``On the Possibility of Superluminal Neutrino Propagation'',
  [arXiv:1109.6296 [hep-ph]].  
 \bibitem{Cacciapaglia:2011ax}
  G.~Cacciapaglia, A.~Deandrea, L.~Panizzi,
  ``Superluminal neutrinos in long baseline experiments and SN1987a'',
  [arXiv:1109.4980 [hep-ph]].
 \bibitem{Bi:2011nd}
  X.~-J.~Bi, P.~-F.~Yin, Z.~-H.~Yu, Q.~Yuan,
  ``Constraints and tests of the OPERA superluminal neutrinos'', 
  [arXiv:1109.6667 [hep-ph]].
 \bibitem{Klinkhamer:2011mf}
  F.~R.~Klinkhamer,
  ``Superluminal muon-neutrino velocity from a Fermi-point-splitting model of Lorentz violation'',
  [arXiv:1109.5671 [hep-ph]].
 \bibitem{Gubser:2011mp}
  S.~S.~Gubser,
  ``Superluminal neutrinos and extra dimensions: Constraints from the null energy condition'',
  Phys.\ Lett.\  {\bf B705 } (2011)  279-281.
  [arXiv:1109.5687 [hep-th]].
 \bibitem{Kehagias:2011cb}
  A.~Kehagias,
  ``Relativistic Superluminal Neutrinos'',
  [arXiv:1109.6312 [hep-ph]].
 \bibitem{Wang:2011sz}
  P.~Wang, H.~Wu, H.~Yang,
  ``Superluminal neutrinos and domain walls'',
  [arXiv:1109.6930 [hep-ph]].
  \bibitem{Saridakis:2011eq}
  E.~N.~Saridakis,
  ``Superluminal neutrinos in Horava-Lifshitz gravity'',
  [arXiv:1110.0697 [gr-qc]].
  \bibitem{Winter:2011zf}
  W.~Winter,
  ``How large is the fraction of superluminal neutrinos at OPERA?'',
  [arXiv:1110.0424 [hep-ph]].
  \bibitem{Alexandre:2011kr}
  J.~Alexandre,
  ``Lifshitz-type Quantum Field Theories in Particle Physics'',
  Int.\ J.\ Mod.\ Phys.\  {\bf A26 } (2011)  4523.
  [arXiv:1109.5629 [hep-ph]].
   \bibitem{Klinkhamer:2011iz}
  F.~R.~Klinkhamer, G.~E.~Volovik,
  ``Superluminal neutrino and spontaneous breaking of Lorentz invariance'',
  Pisma Zh.\ Eksp.\ Teor.\ Fiz.\  {\bf 94 } (2011)  731.
  [arXiv:1109.6624 [hep-ph]].
  \bibitem{Cowsik:2011wv}
  R.~Cowsik, S.~Nussinov, U.~Sarkar,
  ``Superluminal Neutrinos at OPERA Confront Pion Decay Kinematics'',
  [arXiv:1110.0241 [hep-ph]].
  \bibitem{Maccione:2011fr}
  L.~Maccione, S.~Liberati, D.~M.~Mattingly,
  ``Violations of Lorentz invariance in the neutrino sector after OPERA'',
  [arXiv:1110.0783 [hep-ph]].
  \bibitem{Dass:2011yj}
  N.~D.~H.~Dass,
  ``OPERA, SN1987a and energy dependence of superluminal neutrino velocity'',
  [arXiv:1110.0351 [hep-ph]].
  \bibitem{Carmona:2011zg}
  J.~M.~Carmona and J.~L.~Cortes,
  ``Constraints from Neutrino Decay on Superluminal Velocities'',
  arXiv:1110.0430 [hep-ph].
\bibitem{Coleman:1997xq}
  S.~R.~Coleman, S.~L.~Glashow,
  ``Cosmic ray and neutrino tests of special relativity'',
  Phys.\ Lett.\  {\bf B405 } (1997)  249-252.
  [hep-ph/9703240].
\bibitem{Coleman:1998ti}
  S.~R.~Coleman, S.~L.~Glashow,
  ``High-energy tests of Lorentz invariance'',
  Phys.\ Rev.\  {\bf D59 } (1999)  116008.
  [hep-ph/9812418].
\bibitem{Liberati:2001cr}
  S.~Liberati, T.~A.~Jacobson and D.~Mattingly,
  ``High energy constraints on Lorentz symmetry violations'',
  arXiv:hep-ph/0110094.
\bibitem{Jacobson:2001tu}
  T.~Jacobson, S.~Liberati and D.~Mattingly,
  ``TeV astrophysics constraints on Planck scale Lorentz violation'',
  Phys.\ Rev.\  D {\bf 66} (2002) 081302
  [arXiv:hep-ph/0112207].
\bibitem{Jacobson:2002hd}
  T.~Jacobson, S.~Liberati and D.~Mattingly,
  ``Threshold effects and Planck scale Lorentz violation: Combined constraints from high energy astrophysics'',
  Phys.\ Rev.\  D {\bf 67} (2003) 124011
  [arXiv:hep-ph/0209264].
\bibitem{Mattingly:2002ba}
  D.~Mattingly, T.~Jacobson, S.~Liberati,
  ``Threshold configurations in the presence of Lorentz violating dispersion relations'',
  Phys.\ Rev.\  {\bf D67 } (2003)  124012.
  [hep-ph/0211466].
\bibitem{Jacobson:2002ye}
  T.~Jacobson, S.~Liberati and D.~Mattingly,
  ``Lorentz violation and Crab synchrotron emission: A new constraint far beyond the Planck scale'',
  Nature {\bf 424} (2003) 1019
  [arXiv:astro-ph/0212190].
\bibitem{Jacobson:2003ty}
  T.~Jacobson, S.~Liberati and D.~Mattingly,
  ``Comments on `Improved limit on quantum-spacetime modifications of  Lorentz symmetry from observations of gamma-ray blazars' '',
  arXiv:gr-qc/0303001.
\bibitem{Jacobson:2003bn}
  T.~A.~Jacobson, S.~Liberati, D.~Mattingly and F.~W.~Stecker,
  ``New limits on Planck scale Lorentz violation in QED'',
  Phys.\ Rev.\ Lett.\  {\bf 93} (2004) 021101
  [arXiv:astro-ph/0309681].
\bibitem{Jacobson:2004qt}
  T.~Jacobson, S.~Liberati and D.~Mattingly,
  ``Quantum gravity phenomenology and Lorentz violation'',
  Springer Proc.\ Phys.\  {\bf 98} (2005) 83
  [arXiv:gr-qc/0404067].
\bibitem{Jacobson:2004rj}
  T.~Jacobson, S.~Liberati and D.~Mattingly,
  ``Astrophysical bounds on Planck suppressed Lorentz violation'',
  Lect.\ Notes Phys.\  {\bf 669} (2005) 101
  [arXiv:hep-ph/0407370].
\bibitem{Jacobson:2005bg}
  T.~Jacobson, S.~Liberati and D.~Mattingly,
  ``Lorentz violation at high energy: concepts, phenomena and astrophysical constraints'',
  Annals Phys.\  {\bf 321} (2006) 150
  [arXiv:astro-ph/0505267].
\bibitem{Mattingly:2005re}
  D.~Mattingly,
  ``Modern tests of Lorentz invariance'',
  Living Rev.\ Rel.\  {\bf 8 } (2005)  5.
  [gr-qc/0502097].
\bibitem{NBI-HE-78-10}
  H.~B.~Nielsen and M.~Ninomiya,
  ``Beta Function In A Noncovariant Yang-mills Theory,''
  Nucl.\ Phys.\ B\ {\bf 141} (1978) 153.
\bibitem{NBI-HE-82-42}
  S.~Chadha and H.~B.~Nielsen,
  ``Lorentz Invariance As A Low-energy Phenomenon,''
  Nucl.\ Phys.\ B\ {\bf 217} (1983) 125.
\bibitem{NBI-HE-82-30}
  H.~B.~Nielsen and I.~Picek,
  ``Lorentz Noninvariance,''
  Nucl.\ Phys.\ B\ {\bf 211} (1983) 269
   [Addendum-ibid.\ B\ {\bf 242} (1984) 542].
\bibitem{187008}
  H.~B.~Nielsen and I.~Picek,
  ``Lorentz Noninvariance. (addendum) On A Possible Subtraction For The Lorentz Noninvariant Model,''
  Nucl.\ Phys.\ B\ {\bf 242} (1984) 542.
\bibitem{NBI-HE-82-9}
  H.~B.~Nielsen and I.~Picek,
  ``Redei Like Model And Testing Lorentz Invariance,''
  Phys.\ Lett.\ B\ {\bf 114} (1982) 141.
 \bibitem{Colladay:1998fq}
  D.~Colladay and V.~A.~Kostelecky,
  ``Lorentz-violating extension of the standard model'',
  Phys.\ Rev.\  D {\bf 58} (1998) 116002
  [arXiv:hep-ph/9809521].
\bibitem{Kostelecky:1988zi}
  V.~A.~Kostelecky and S.~Samuel,
  ``Spontaneous Breaking of Lorentz Symmetry in String Theory'',
  Phys.\ Rev.\  D {\bf 39} (1989) 683.
\bibitem{Kostelecky:2003fs}
  V.~A.~Kostelecky,
  ``Gravity, Lorentz violation, and the standard model'',
  Phys.\ Rev.\  D {\bf 69} (2004) 105009
  [arXiv:hep-th/0312310].
  \bibitem{Kostelecky:2000mm}
  V.~A.~Kostelecky and R.~Lehnert,
  ``Stability, causality, and Lorentz and CPT violation'',
  Phys.\ Rev.\  D {\bf 63} (2001) 065008
  [arXiv:hep-th/0012060].
\bibitem{Kostelecky:2002hh}
  V.~A.~Kostelecky and M.~Mewes,
  ``Signals for Lorentz violation in electrodynamics'',
  Phys.\ Rev.\  D {\bf 66} (2002) 056005
  [arXiv:hep-ph/0205211].
  \bibitem{Kostelecky:2003cr}
  V.~A.~Kostelecky and M.~Mewes,
  ``Lorentz and CPT violation in neutrinos'',
  Phys.\ Rev.\  D {\bf 69} (2004) 016005
  [arXiv:hep-ph/0309025].
  \bibitem{Kostelecky:1999mr}
  V.~A.~Kostelecky and C.~D.~Lane,
  ``Constraints on Lorentz violation from clock-comparison experiments'',
  Phys.\ Rev.\  D {\bf 60} (1999) 116010
  [arXiv:hep-ph/9908504].
\bibitem{Kostelecky:2001mb}
  V.~A.~Kostelecky and M.~Mewes,
  ``Cosmological constraints on Lorentz violation in electrodynamics'',
  Phys.\ Rev.\ Lett.\  {\bf 87} (2001) 251304
  [arXiv:hep-ph/0111026].
  \bibitem{Bear:2000cd}
  D.~Bear, R.~E.~Stoner, R.~L.~Walsworth, V.~A.~Kostelecky and C.~D.~Lane,
  ``Limit on Lorentz and CPT violation of the neutron using a two-species noble-gas maser'',
  Phys.\ Rev.\ Lett.\  {\bf 85} (2000) 5038
  [Erratum-ibid.\  {\bf 89} (2002) 209902]
  [arXiv:physics/0007049].
 \bibitem{Anselmi:2011zz}
  D.~Anselmi,
  ``Renormalization And Lorentz Symmetry Violation'',
  PoS C {\bf LAQG08} (2011) 010.
  \bibitem{Anselmi:2011bp}
  D.~Anselmi and D.~Buttazzo,
  ``Distance Between Quantum Field Theories As A Measure Of Lorentz Violation'',
  Phys.\ Rev.\  D {\bf 84} (2011) 036012
  [arXiv:1105.4209 [hep-ph]].
 \bibitem{Anselmi:2007zz}
  D.~Anselmi,
 ``Renormalization of Lorentz violating theories'',
{\it Prepared for 4th Meeting on CPT and Lorentz Symmetry, Bloomington, Indiana, 8-11 Aug 2007}
\bibitem{Anselmi:2011ae}
  D.~Anselmi and M.~Taiuti,
  ``Vacuum Cherenkov Radiation In Quantum Electrodynamics With High-Energy Lorentz Violation'',
  Phys.\ Rev.\  D {\bf 83} (2011) 056010
  [arXiv:1101.2019 [hep-ph]].
 \bibitem{Anselmi:2011ac}
  D.~Anselmi and E.~Ciuffoli,
  ``Low-energy Phenomenology Of Scalarless Standard-Model Extensions With High-Energy Lorentz Violation'',
  Phys.\ Rev.\  D {\bf 83} (2011) 056005
  [arXiv:1101.2014 [hep-ph]].
  \bibitem{Anselmi:2010zh}
  D.~Anselmi and E.~Ciuffoli,
  ``Renormalization Of High-Energy Lorentz Violating Four Fermion Models'',
  Phys.\ Rev.\  D {\bf 81} (2010) 085043
  [arXiv:1002.2704 [hep-ph]].
\bibitem{Anselmi:2009ng}
  D.~Anselmi and M.~Taiuti,
  ``Renormalization Of High-Energy Lorentz Violating QED'',
  Phys.\ Rev.\  D {\bf 81} (2010) 085042
  [arXiv:0912.0113 [hep-ph]].
  \bibitem{Anselmi:2009vz}
  D.~Anselmi,
  ``Standard Model Without Elementary Scalars And High Energy Lorentz Violation'',
  Eur.\ Phys.\ J.\  C {\bf 65} (2010) 523
  [arXiv:0904.1849 [hep-ph]].
 \bibitem{Anselmi:2008bt}
  D.~Anselmi,
  ``Weighted power counting, neutrino masses and Lorentz violating extensions of the Standard Model'',
  Phys.\ Rev.\  D {\bf 79} (2009) 025017
  [arXiv:0808.3475 [hep-ph]].
 \bibitem{Anselmi:2008bs}
  D.~Anselmi,
  ``Weighted power counting and Lorentz violating gauge theories. II: Classification'',
  Annals Phys.\  {\bf 324} (2009) 1058
  [arXiv:0808.3474 [hep-th]].
 \bibitem{Anselmi:2008bq}
  D.~Anselmi,
  ``Weighted power counting and Lorentz violating gauge theories. I: General properties'',
  Annals Phys.\  {\bf 324} (2009) 874
  [arXiv:0808.3470 [hep-th]].
  \bibitem{Anselmi:2008ry}
  D.~Anselmi,
  ``Weighted scale invariant quantum field theories'',
  JHEP {\bf 0802} (2008) 051
  [arXiv:0801.1216 [hep-th]].
\bibitem{Anselmi:2007ri}
  D.~Anselmi and M.~Halat,
  ``Renormalization of Lorentz violating theories'',
  Phys.\ Rev.\  D {\bf 76} (2007) 125011
  [arXiv:0707.2480 [hep-th]].
\bibitem{Horava:2009uw}
  P.~Horava,
  ``Quantum Gravity at a Lifshitz Point'',
  Phys.\ Rev.\  {\bf D79 } (2009)  084008.
  [arXiv:0901.3775 [hep-th]].
\bibitem{Visser:2009fg}
  M.~Visser,
  ``Lorentz symmetry breaking as a quantum field theory regulator'',
  Phys.\ Rev.\  {\bf D80 } (2009)  025011.
  [arXiv:0902.0590 [hep-th]].
\bibitem{Visser:2009ys}
  M.~Visser,
  ``Power-counting renormalizability of generalized Horava gravity'',
  [arXiv:0912.4757 [hep-th]].
\bibitem{Sotiriou:2009bx}
  T.~P.~Sotiriou, M.~Visser, S.~Weinfurtner,
  ``Quantum gravity without Lorentz invariance'',
  JHEP {\bf 0910 } (2009)  033.
  [arXiv:0905.2798 [hep-th]].
\bibitem{Sotiriou:2009gy}
  T.~P.~Sotiriou, M.~Visser, S.~Weinfurtner,
  ``Phenomenologically viable Lorentz-violating quantum gravity'',
  Phys.\ Rev.\ Lett.\  {\bf 102 } (2009)  251601.
  [arXiv:0904.4464 [hep-th]].
\bibitem{Weinfurtner:2010hz}
  S.~Weinfurtner, T.~P.~Sotiriou, M.~Visser,
  ``Projectable Horava-Lifshitz gravity in a nutshell'',
  J.\ Phys.\ Conf.\ Ser.\  {\bf 222 } (2010)  012054.
  [arXiv:1002.0308 [gr-qc]].
 \bibitem{Visser:2011mf}
  M.~Visser,
  ``Status of Horava gravity: A personal perspective'',
  J.\ Phys.\ Conf.\ Ser.\  {\bf 314}, 012002 (2011).
  [arXiv:1103.5587 [hep-th]].
\bibitem{Judes:2002bw}
  S.~Judes, M.~Visser,
  ``Conservation laws in `Doubly special relativity' '',
  Phys.\ Rev.\  {\bf D68 } (2003)  045001.
  [arXiv:gr-qc/0205067 [gr-qc]].
\bibitem{Liberati:2004ju}
  S.~Liberati, S.~Sonego, M.~Visser,
  ``Interpreting doubly special relativity as a modified theory of measurement'',
  Phys.\ Rev.\  {\bf D71 } (2005)  045001.
  [gr-qc/0410113].
  \bibitem{Barcelo:2005fc}
  C.~Barcelo, S.~Liberati, M.~Visser,
  ``Analogue gravity'',
  Living Rev.\ Rel.\  {\bf 8 } (2005)  12.
  [gr-qc/0505065].
  \enlargethispage{35pt}
  \bibitem{Visser:1997ux}
  M.~Visser,
  ``Acoustic black holes: Horizons, ergospheres, and Hawking radiation'',
  Class.\ Quant.\ Grav.\  {\bf 15}, 1767-1791 (1998).
  [gr-qc/9712010].
\end{thebibliography}
\end{document}